\documentclass[final,3p,times]{elsarticle}
\usepackage{amssymb}
\usepackage{graphicx,setspace}
\usepackage{psfrag}
\usepackage{amsfonts}
\expandafter\let\csname equation*\endcsname\relax
\expandafter\let\csname endequation*\endcsname\relax
\usepackage{amsmath}
\usepackage{amssymb, amsthm}
\usepackage{mathrsfs}
\usepackage{epstopdf}
\usepackage{float}
\usepackage{lineno,hyperref}
\usepackage{color}
\usepackage{subfigure}
\usepackage{amsmath}
\usepackage{dsfont} % it is the package of using the "\mathds{} "
\usepackage{amssymb} % it is the package of using the "\mathbb{}"
\usepackage{graphicx,color}
\usepackage{extarrows}

\usepackage{graphicx}

\usepackage{epstopdf}
\journal{arXiv}

\begin{document}
\newtheorem{definition}{Definition}[section]
\newtheorem{lemma}{Lemma}[section]
\newtheorem{remark}{Remark}[section]
\newtheorem{theorem}{Theorem}[section]
\newtheorem{proposition}{Proposition}
\newtheorem{assumption}{Assumption}
\newtheorem{example}{Example}
\newtheorem{corollary}{Corollary}[section]
\def\ep{\varepsilon}
\def\Rn{\mathbb{R}^{n}}
\def\Rm{\mathbb{R}^{m}}
\def\E{\mathbb{E}}
\def\hte{\hat\theta}
%\numberwithin{theorem}{section}
%\numberwithin{definition}{section}
\renewcommand{\theequation}{\thesection.\arabic{equation}}
\begin{frontmatter}



%\title{Searching zero-temperature transition paths in stochastic superconducting tunnel junction systems}
\title{Stochastic dynamics of the resistively shunted superconducting tunnel junction system under the impact of thermal fluctuations}
%Most probable transport paths
\author{Shenglan Yuan\corref{cor1}\fnref{addr1,addr2}}\ead{shenglanyuan@gbu.edu.cn}\cortext[cor1]{Corresponding author}
\begin{center}
%{\small \textbf{Salute to Professor Francis Comets and may his soul rest in peace}}
\end{center}

\address[addr1]{\rm Department of Mathematics, School of Sciences, Great Bay University, Dongguan 523000, China }
\address[addr2]{\rm Great Bay Institute for Advanced Study, Songshan Lake International Innovation
Entrepreneurship Community A5, Dongguan 523000, China }
%\author{Peter H\"{a}nggi\fnref{addr2}}
%\ead{hanggi@physik.uni-augsburg.de}
%\address[addr2]{\rm Institut f$\rm\ddot{u}$r Mathematik, Universit$\rm\ddot{a}$t Augsburg,
%86135, Augsburg, Germany}
%\address[addr3]{\rm Institut f\"{u}r Physik, Universit\"{a}t Augsburg, 86135 Augsburg, Germany}

\begin{abstract}
In this work, a Josephson junction consisting of two superconducting layers
sandwiching an insulating layer is explored, which is subject to the effects of thermal fluctuations. The precise expressions for the evolution of Josephson phase and the supercurrent through the junction are derived.  A clockwise hysteresis cycle in the current-voltage characteristic curve  is demonstrated mathematically. Additionally,  the bifurcation of a planar limit cycle is established.
 The numerous stochastic thermodynamic properties of the resistively shunted superconducting tunnel junction system are described, considering the influence of three specific parameters:
the conductance, the current bias and the noise intensity. Moreover, the probability density is characterized using the Fokker-Planck equation.
\end{abstract}

\begin{keyword}
Stochastic dynamics;  Resistively shunted junction; Thermal fluctuations; Hysteresis; Bifurcation



\emph{2020 Mathematics Subject Classification}: 82D55, 82B26, 82B30 .

\end{keyword}

\end{frontmatter}
\section{Introduction}

The Josephson effect \cite{J} is a phenomenon that occurs when two superconductors are placed in proximity, with some barrier or restriction between them. This effect produces a current known as a supercurrent, which flows continuously without any applied voltage across a device called a Josephson junction. These junctions consist of two or more superconductors coupled by a weak link. The weak link can be a thin insulating barrier or another type of restriction \cite{ROWSBS}.

The superconducting tunnel junction is a type of Josephson junction that takes
advantage of quantum tunneling and superconductivity to create the Josephson effect \cite{BZLDFS}.
It is an electronic device consisting of two superconductors, such as niobium, tantalum, or hafnium, separated by a very thin layer of insulating material, such as aluminium oxide. It is also known as a superconductor-insulator-superconductor tunnel junction, where a barrier can be created by separating two conductors with an insulator.

To fabricate a superconducting tunnel junction on an insulating substrate
such as silicon, sapphire, or magnesium fluoride, the first layer of a thin film of
superconducting metal with a typical thickness of several nanometers is deposited inside a vacuum chamber. This layer is then oxidized to form a high-quality insulating tunnel barrier after oxygen gas is introduced into the chamber. The final layer of superconducting metal
is overlapped with the top of the previous layers when the vacuum is restored.

The superconducting tunnel junction has a wide range of practical applications, including high-sensitivity radio astronomy \cite{W}, single-photon detection, superconducting quantum interference devices (SQUIDs) \cite{SKK,ZBSH}, quantum computers \cite{BR}, rapid single flux quantum (RSFQ) fast logic circuits, and Josephson voltage standards \cite{SLFMJ}. It is also used in high-sensitivity detectors of electromagnetic radiation, magnetometers, high-speed digital circuit elements, and quantum computing circuits \cite{RWB,TBZ}. Additionally, it finds important applications in precision measurements of voltages and magnetic fields, multijunction solar cells, quantum-mechanical circuits, superconducting qubits, as well as RSFQ digital electronics \cite{BK,KBLS}.

However, superconducting tunnel junction devices are open systems that interact with one or more fluctuating environments \cite{SL}.
Quantum noise is usually observed during experiments, which is typically unwanted but inescapable due to factors such as thermal fluctuations,
mechanical vibrations, industrial influences of voltage from power supplies, measurement operations \cite{OUB}.
Even though quantum noise is an uncontrolled variation deviating from a desired value, an accurate description of the superconducting tunnel junction device performance should seriously take into account these random effects \cite{T}.

The resistively shunted superconducting tunnel junction  system is an important building block of
superconducting electronic circuits \cite{CIKZ,M}. It is a fundamental component characterized by a critical current that flows through the junction without resistance.
In spite of this, when a voltage bias is applied across the junction, the system enters a resistive state because of  the tunneling of Cooper pairs from one superconductor to the other.
This resistive state is accompanied by the generation of heat, which in turn leads to the influence of thermal fluctuations on the system's dynamics \cite{BST,NRNKK}.

Stochastic thermodynamics particularly refers to a system's behavior that is characterized by random fluctuations caused by the heating effect of
thermal noise. In the resistively shunted superconducting tunnel junction system, these fluctuations result in a
time-varying voltage and current across the junction, making its behavior unpredictable.
The stochastic dynamics of this noise-affected system reveal that these fluctuations could significantly impact the system's behavior \cite{KMDG}. Specifically, the thermal fluctuations could influence the switching behavior of the junction, affecting its transition between superconducting and resistive states. Additionally, these fluctuations also affect the voltage-current characteristics of the junction, leading to deviations from the ideal behavior predicted by theoretical models.
Understanding and modeling
this stochastic behavior is crucial for optimizing the performance of superconducting electronic circuits \cite{WT}.

The studies of stochastic dynamics in the resistively shunted superconducting tunnel junction system in the presence of thermal noise constitute an active area of research.
McCumber \cite{M} demonstrated that the AC (Alternating Current) impedance observed by the junction has an impact on  both the response time to changes in bias and the DC (Direct Current) voltage-current characteristics.
Kautz \cite{K} designed zero-bias Josephson voltage standards by utilizing superconducting tunnel junctions as the core component and  calculated the quasipotential difference that represents the likelihood of escape from a basin of attraction for nonequilibrium systems at low temperatures.
Spiechowicz, H\"{a}nggi and J. {\L}uczka \cite{SHL} investigated the stochastic dynamics  in an asymmetric SQUID composed of a loop containing three Josephson junctions, identified specific parameter sets where the ratchet effect is most pronounced, and showed how the direction of transport can be precisely controlled by adjusting the external magnetic flux.
Guarcello, Bergeret and Citro \cite{GBC} conducted an investigation into the switching current distributions exhibited by ferromagnetic anomalous Josephson junctions while subjecting them to a linearly increasing bias current. Their study analyzed the intricate interplay among noise, magnetization, phase dynamics, and the statistical characteristics of the switching current distribution.

The aim of this research is to understand the stochastic dynamics of the resistively shunted superconducting tunnel junction system under the external excitation of thermal fluctuations. This knowledge can aid in the development of more accurate models for predicting the system's behavior, as well as exploring new techniques for controlling and mitigating the impact of thermal fluctuations on circuit performance.

This work is arranged as follows. In Section \ref{IV}, the Josephson equations of supercurrent and voltage are derived by employing the wave function of the superconducting tunnel junction and the time-dependent Schr\"{o}dinger  equation it satisfies.
In Section \ref{DM}, the expression for the total junction current is explored according to Kirchhoff's current law, making it dimensionless. The fluctuation-dissipation theorem is then applied to introduce noise, thereby obtaining the resistively shunted superconducting tunnel junction system driven by thermal noise.
In Section \ref{DRSTJ}, it is found that the resistively shunted superconducting tunnel junction system corresponds to a Hamiltonian system in the absence of parameters. The energy surface and phase portrait are simulated. When the parameters are non-zero, a clockwise hysteresis is observed in the typical I-V characteristic curve of the system, and a bifurcation of a planar limit cycle occurs. Most importantly, the stochastic dynamics of the system influenced by thermal fluctuation are depicted through numerical simulations.
In Section \ref{FPE}, the Fokker-Planck equation for the resistively shunted superconducting tunnel junction system under the impact of thermal fluctuations is calculated, and the probability density is simulated based on this equation for certain parameter settings.
In Section \ref{SO}, the findings are summarized and several future threads of discussion are identified.

\section{Josephson equations}\label{IV}
\begin{figure}[H]
\vspace*{1pt}
\begin{center}
\begin{minipage}{2in}
		\includegraphics[width=2in]{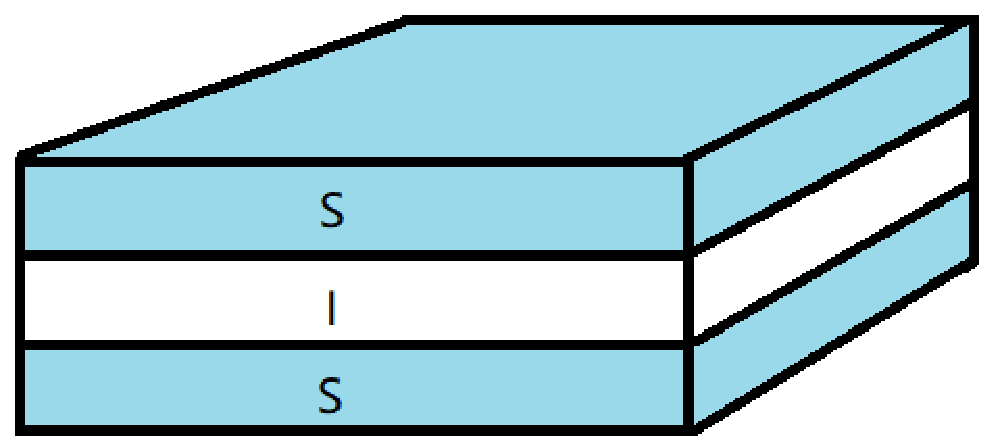}
	\end{minipage}
\end{center}
\caption{A SIS junction consists of two bulk superconductors separated by a thin insulator layer.}\label{bulk}	
\end{figure}

Referring to Figure \ref{bulk}, the basic Josephson junction (or SIS junction) is built from two bulk superconductors, separated by a very thin layer of an insulator. The separation layer is of the order of few angstroms to few nanometers.
When the superconductor's temperature $T$ is cooled below its critical temperature $T_c$, the electrons within it begin to pair up, forming what are known as cooper pairs. These electron pairs have a boson nature, allowing all the electrons in the superconductor to condense into the ground state. This condensation  allows the electrons  in the superconductor to be treated as a single quantum particle, described by a macroscopic wave function that possesses a well-defined  quantum phase:
\begin{equation}\label{qphase}
\Psi=\Psi_0e^{i\varphi}.
\end{equation}
The wave functions of upper and lower superconductors are:
\begin{equation}\label{UL}
\begin{array}{ll}
|U\rangle=\sqrt{\rho_U}e^{i\varphi_U}, \\
|L\rangle=\sqrt{\rho_L}e^{i\varphi_L},
\end{array}
\end{equation}
where $\rho_U=\rho_L=:\rho>0$ is assumed to be identical amplitude, which is the cooper pair density in both upper and lower superconducting bulks.
The total Hamiltonian $H:=H_U+H_L+H_M$ consists of three terms including the Hamiltonians of the upper and lower superconducting bulks and a mix of them together:
\begin{equation}\label{HULM}
\begin{array}{ll}
H_U=E_U|U\rangle\langle U|,  \\
H_L=E_L|L\rangle\langle L|,  \\
H_M=K(|U\rangle\langle L|+|L\rangle\langle U|),
\end{array}
\end{equation}
where $E_U=eV$ and $E_L=-eV$ are the energy levels in the two superconductors, and the constant $K$ is a characteristic of the junction. Notice that the energy difference $E_U-E_L$
between the upper and lower superconductors is $2eV$.

The time-dependent Schr\"{o}dinger equation $i\hbar\dot{\Psi}=H\Psi$ for the states $|U\rangle$ and $|L\rangle$  is:
\begin{equation}\label{SUL}
\left\{
\begin{array}{ll}
i\hbar\frac{d|U\rangle}{d\tau}=E_{U}|U\rangle+K|L\rangle, \\[0.3ex]
i\hbar\frac{d|L\rangle}{d\tau}=E_{L}|L\rangle+K|U\rangle,
\end{array}
\right.
\end{equation}
where $\hbar= h/2\pi$, and  $h$ is the Planck constant.

In order to solve the Schr\"{o}dinger equation \eqref{SUL}, we first calculate the time
derivative of the order parameter in the upper superconductor:
\begin{equation*}
\frac{d|U\rangle}{d\tau}=\frac{d\sqrt{\rho_U}}{d\tau}e^{i\varphi_U}+\sqrt{\rho_U}\Big(i\frac{d\varphi_U}{d\tau}e^{i\varphi_U}\Big)=\Big(\frac{d\sqrt{\rho_U}}{d\tau}+i\sqrt{\rho_U}\frac{d\varphi_U}{d\tau}\Big)e^{i\varphi_U},
\end{equation*}
and therefore
\begin{equation}\label{dUdt}
\Big(\frac{d\sqrt{\rho_U}}{d\tau}+i\sqrt{\rho_U}\frac{d\varphi_U}{d\tau}\Big)e^{i\varphi_U}=\frac{1}{i\hbar}\big(E_{U}\sqrt{\rho_U}e^{i\varphi_U}+K\sqrt{\rho_L}e^{i\varphi_L}\big).
\end{equation}
The quantum phase difference between the  upper and lower superconductor bulks is called the Josephson
phase: $$\phi=\varphi_L-\varphi_U.$$
The equation \eqref{dUdt} can therefore be rewritten as:
\begin{equation*}
\frac{d\sqrt{\rho_U}}{d\tau}+i\sqrt{\rho_U}\frac{d\varphi_U}{d\tau}=\frac{1}{i\hbar}\big(E_{U}\sqrt{\rho_U}+K\sqrt{\rho_L}e^{i\phi}\big),
\end{equation*}
and its complex conjugate equation is
\begin{equation*}
\frac{d\sqrt{\rho_U}}{d\tau}-i\sqrt{\rho_U}\frac{d\varphi_U}{d\tau}=\frac{1}{-i\hbar}\big(E_{U}\sqrt{\rho_U}+K\sqrt{\rho_L}e^{-i\phi}\big).
\end{equation*}
Add the above two conjugate equations together to eliminate $\frac{d\varphi_U}{d\tau}$:
\begin{equation*}
2\frac{d\sqrt{\rho_U}}{d\tau}=\frac{K\sqrt{\rho_L}}{i\hbar}(e^{i\phi}-e^{-i\phi})=\frac{K\sqrt{\rho_L}}{\hbar}2\sin(\phi).
\end{equation*}
Because $$\frac{d\sqrt{\rho_U}}{d\tau}=\frac{1}{2\sqrt{\rho_U}}\frac{d\rho_U}{d\tau},$$ we obtain
\begin{equation*}
\frac{d\rho_U}{d\tau}=\frac{2K\sqrt{\rho_U\rho_L}}{\hbar}\sin(\phi)=\frac{2K\rho}{\hbar}\sin(\phi).
\end{equation*}
Similarly, we subtract the two conjugate equations to eliminate $\frac{d\sqrt{\rho_U}}{d\tau}$:
\begin{equation*}
2i\sqrt{\rho_U}\frac{d\varphi_U}{d\tau}=\frac{1}{i\hbar}\big[2E_{U}\sqrt{\rho_U}+K\sqrt{\rho_L}\big(e^{i\phi}+e^{-i\phi}\big)\big],
\end{equation*}
which yields
\begin{equation*}
\frac{d\varphi_U}{d\tau}=-\frac{1}{\hbar}\big(E_{U}+K\sqrt{\frac{\rho_L}{\rho_U}}\cos(\phi)\big)=-\frac{1}{\hbar}\big(E_{U}+K\cos(\phi)\big).
\end{equation*}

Now, for the lower superconductor we can compute the time
derivative
\begin{equation*}
\frac{d|L\rangle}{d\tau}=\frac{d\sqrt{\rho_L}}{d\tau}e^{i\varphi_L}+\sqrt{\rho_L}\Big(i\frac{d\varphi_L}{d\tau}e^{i\varphi_L}\Big)=\Big(\frac{d\sqrt{\rho_L}}{d\tau}+i\sqrt{\rho_L}\frac{d\varphi_L}{d\tau}\Big)e^{i\varphi_L},
\end{equation*}
which gives
\begin{equation*}
\Big(\frac{d\sqrt{\rho_L}}{d\tau}+i\sqrt{\rho_L}\frac{d\varphi_L}{d\tau}\Big)e^{i\varphi_L}=\frac{1}{i\hbar}\big(E_{L}\sqrt{\rho_L}e^{i\varphi_L}+K\sqrt{\rho_U}e^{i\varphi_U}\big).
\end{equation*}
Since  the difference in phase between the macroscopic wavefunctions that describe the
paired electrons in the two electrodes is  $\phi$, we have
\begin{equation}\label{dLdt}
\frac{d\sqrt{\rho_L}}{d\tau}+i\sqrt{\rho_L}\frac{d\varphi_L}{d\tau}=\frac{1}{i\hbar}\big(E_{L}\sqrt{\rho_L}+K\sqrt{\rho_U}e^{i\phi}\big).
\end{equation}
Separating the equation \eqref{dLdt} into real and imaginary parts, we get
\begin{equation}\label{dpho}
\frac{d\sqrt{\rho_L}}{d\tau}=\frac{K\sqrt{\rho_U}}{\hbar}\sin(\phi)
\end{equation}
and
\begin{equation*}
\frac{d\varphi_L}{d\tau}=-\frac{1}{\hbar}\big(E_{L}+K\cos(\phi)\big).
\end{equation*}
Substituting  $\frac{d\sqrt{\rho_L}}{d\tau}=\frac{1}{2\sqrt{\rho_L}}\frac{d\rho_L}{d\tau}$  into  \eqref{dpho}, we gain
\begin{equation*}
\frac{d\rho_L}{d\tau}=\frac{2K\rho}{\hbar}\sin(\phi).
\end{equation*}

Noting that the evolution of Josephson phase is
\begin{equation*}
\frac{d\phi}{d\tau}=\frac{d\varphi_L}{d\tau}-\frac{d\varphi_U}{d\tau}=-\frac{1}{\hbar}(E_L-E_U)=\frac{2eV}{\hbar}.
\end{equation*}
 This is actually the time derivative
of the quantum phase difference on the junction.

The two Josephson equations are \begin{align}\label{supercurrent}
I&=I_c\sin(\phi),\\ \label{pdifference}
V&=\frac{\hbar}{2e}\frac{d\phi}{d\tau},
\end{align}
where $I$ is the supercurrent through the junction, $I_c:=\frac{2AK\rho}{\hbar}$ is a junction parameter called the critical current for which $A$ is a geometrical factor depended on the area of the junction, and $V$ is the voltage across the junction.
These equations establish a relationship between the supercurrent $I$ and
the voltage $V$ with the phase $\phi$ serving as an intermediate variable.
\begin{remark}
We connect phase of the wave functions to current and voltage across weak link by the equations \eqref{supercurrent}-\eqref{pdifference}.
The Josephson equations indicate that a current $I$ of
magnitude less than $I_c$ can flow without a voltage (DC Josephson effect), and
that the current oscillates with a particular angular frequency $\omega_{J}:=2eV/\hbar$, called the Josephson
frequency, when a constant voltage is present (AC Josephson effect).
\end{remark}

\section{Dimensionless model}\label{DM}
\begin{figure}
\begin{center}
\begin{minipage}{3in}
\leftline{(a)}
\includegraphics[width=3in]{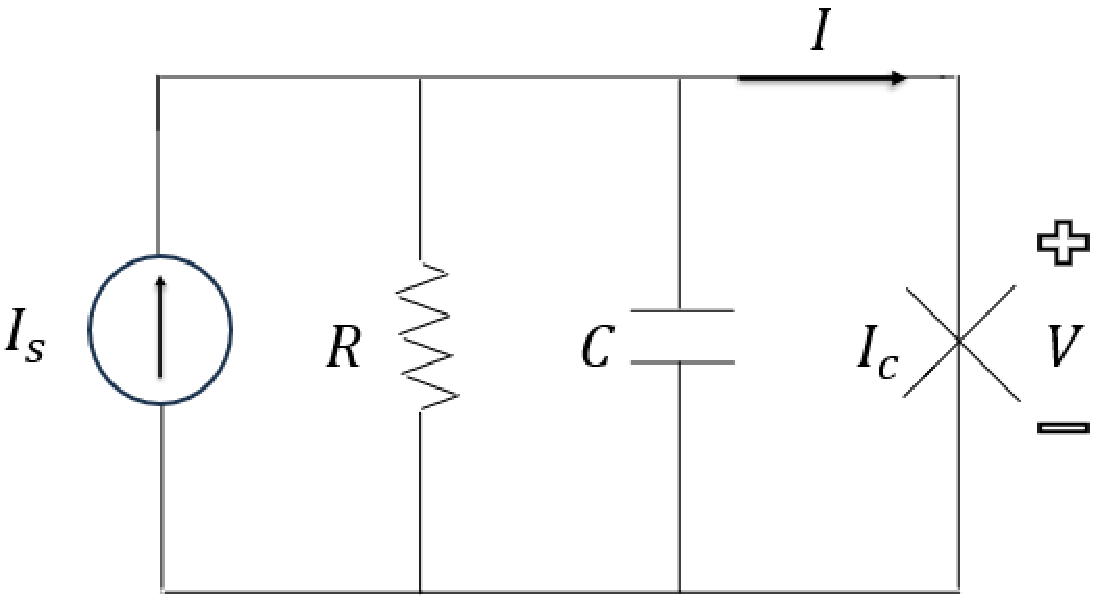}
\end{minipage}
\hfill
  \begin{minipage}{3in}
\leftline{(b)}
\includegraphics[width=3in]{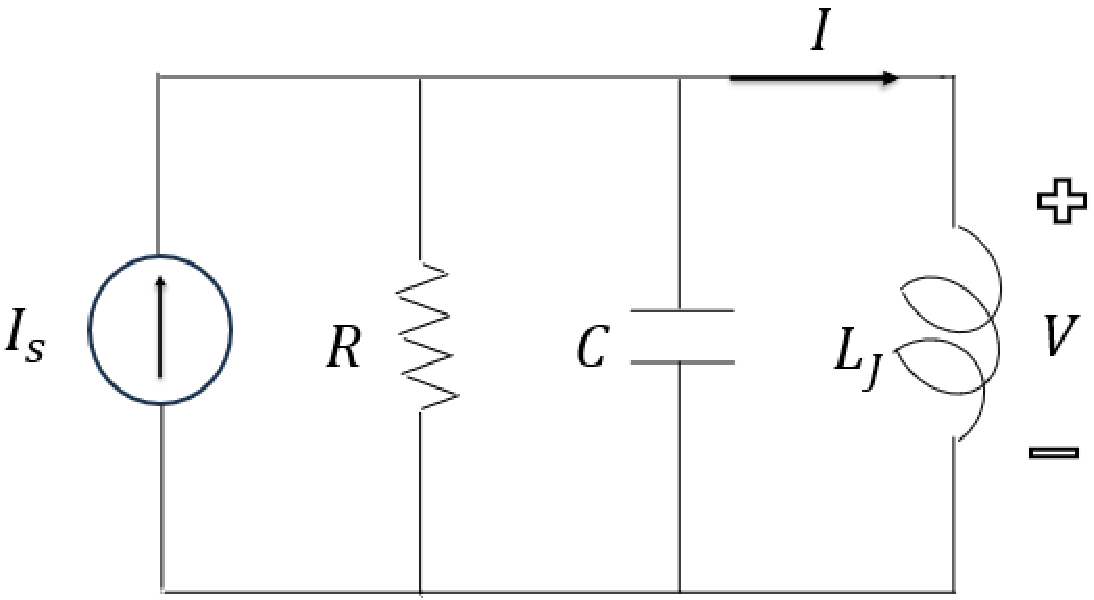}
\end{minipage}
\caption{(a) Full scheme of the experiment.  The junction comprises three parallel lumped elements: a capacitor representing the internal capacitance
of the junction, a shunt resistor, and a Josephson element having a current according to \eqref{supercurrent}. The total junction current $I_S$ is a standard
bias current coming from a source in parallel to the Josephson
junction; (b) When $\phi\ll1$, the linear approximation $\sin(\phi)\approx\phi$ is valid, which means that $\sin(\phi)$ and $\phi$ are very close in this range.
The system of motion, which consists of a single Josephson
junction embedded in a superconducting loop, can be described by the equivalent circuit where the self inductance is $L_J$.}\label{circuit}
\end{center}
\end{figure}

According to Th\'evenin's theorem, the AC impedance of the junction can be represented by a capacitor and a shunt resistor placed in parallel with the Josephson junction that is embedded in a superconducting closed loop, as indicated in Figure \ref{circuit} (a).
Based on Ohm's law in electrical circuit theory that describes the relationship between the current, voltage, and resistance, the normal current through the resistor is expressed mathematically as $\frac{V}{R}$, where $R$ accounts for the resistance.
The  displacement current through the capacitor is $C\frac{dV}{d\tau}$, where $C$ is the effective capacitance of the junction. In accordance with Kirchhoff's current law, the complete expression for the total junction current $I_S$ is
\begin{equation}\label{Kirchhoff}
I_S=C\frac{dV}{d\tau}+\frac{V}{R}+I=\frac{C\hbar}{2e}\frac{d^{2}\phi}{d\tau^{2}}+\frac{\hbar}{2eR}\frac{d\phi}{d\tau}+I_c\sin(\phi),
\end{equation}
by utilizing the relation between the voltage and phase difference as in the equation \eqref{pdifference}.

Further insight is gained from a mechanical analogue of the Josephson junction, which provides an intuitive understanding of its nonlinear dynamics. This analogue is based on the fact that the ideal Josephson element acts as an energy storage device. We make reasonable use of the equations \eqref{supercurrent}-\eqref{pdifference} to derive that the stored energy $U$, as a function of $\phi$, is
\begin{equation*}
U(\phi)=\int_{0}^{\tau} IVd\tau=\int_{0}^{\tau}I_c\sin(\phi)\frac{\hbar}{2e}\frac{d\phi}{d\tau}d\tau=\frac{I_c\hbar}{2e}\int_{\phi_0}^{\phi}\sin(\phi)d\phi=\frac{I_c\hbar}{2e}\big(1-\cos(\phi)\big),
\end{equation*}
where $E_J:=\frac{I_c\hbar}{2e}$ is the Josephson coupling energy, and the reference energy level is set to be at $\phi_0=0$.
The system's nonlinearity arises from the sinusoidal nature of the potential, and the degree of nonlinearity is estimated by the Josephson coupling energy $E_J$, which determines the depth of the oscillations in potential energy.

If we interpret $\phi$ as a position coordinate, then the equation
\eqref{Kirchhoff} can be interpreted as Newton's equation for the motion of a particle in the potential $U(\phi)$ in the presence of damping and external applied forces.
The acceleration of particle is as follows:
\begin{equation}\label{Newtonseq}
\frac{C\hbar}{2e}\frac{d^{2}\phi}{d\tau^{2}}=-I_c\sin(\phi)-\frac{\hbar}{2eR}\frac{d\phi}{d\tau}+I_S=-\frac{2e}{\hbar}U'(\phi)-\frac{\hbar}{2eR}\frac{d\phi}{d\tau}+I_S,
\end{equation}
where
\begin{equation*}
U'(\phi)=\frac{dU}{d\phi}=\frac{I_c\hbar}{2e}\sin(\phi)
\end{equation*}
is the derivative of the potential $U(\phi)$ with respect to $\phi$. Therefore, the dynamical behavior of a Josephson junction is exactly analogous to the motion of a particle in a sinusoidal potential, often referred to as a washboard potential.
The three elements on the right-hand side of equation \eqref{Newtonseq} indicate the components contributing to the force. These components include the force resulting from the negative derivative $-U'(\phi)$ of the potential, a damping force that is proportional to the velocity $d\phi/d\tau$, and an externally applied force with a constant value. In the context of mechanics, the velocity of a particle represents the junction voltage, the viscous damping is equivalent to the junction conductance
$G=\frac{1}{R}$, and the external force $I_S$ represents the current bias. Furthermore, the mass of the particle in the mechanical analogy is very similar to the junction capacitance on the left-hand side of equation \eqref{Newtonseq}.
%\begin{remark}
%The nonlinearity of the system is due to the sinusoidal shape of the potential, and the strength of the nonlinearity is determined by the Josephson coupling
%energy $E_J$, which sets the depth of the potential-energy oscillations.
%\end{remark}

Regarding the case  $\phi\ll1$ in which we have
$\sin(\phi)\approx\phi$ that is linear.
This approximation can be applied to simplify the calculations or analyze the properties of functions. Specifically in the case of $\phi\ll1$,
\begin{equation*}
I=I_c\sin(\phi)\overset{\sin(\phi)\approx\phi}{\Longrightarrow}I=I_c\phi\Longrightarrow\frac{d\phi}{d\tau}=\frac{1}{I_c}\frac{dI}{d\tau}\overset{V=\frac{\hbar}{2e}\frac{d\phi}{d\tau}}{\Longrightarrow}V=\frac{\hbar}{2eI_c}\frac{dI}{dt},
 \end{equation*}
where the ideal element  $L_J:=\frac{\hbar}{2eI_c}$ is the Josephson inductance of an inductor. It is related to the Josephson coupling energy by $E_J=L_JI_c^2$.

Due to the equations \eqref{supercurrent}-\eqref{pdifference}, we apply the chain rule to calculate the time derivative of the supercurrent:
\begin{equation*}
\frac{dI}{d\tau}=\frac{dI}{d\phi}\frac{d\phi}{d\tau}=I_c\cos(\phi)\frac{2e}{\hbar},
\end{equation*}
which can be rearranged in the form of the current-voltage characteristic of an inductor:
\begin{equation*}
V=\frac{\hbar}{2eI_c\cos(\phi)}\frac{dI}{d\tau},
\end{equation*}
where
\begin{equation*}
L(\phi):=\frac{\hbar}{2eI_c\cos(\phi)}=\frac{L_J}{\cos(\phi)}
\end{equation*}
provides the useful expression for the kinetic inductance as a function of the Josephson phase. It follows that $L(0)=L_J$.
In essence, the kinetic behavior of the Josephson junction is similar to that of an inductor, stemming from the kinetic energy of the charge carriers. In Figure \ref{circuit} (b), the loop can be replaced by an equivalent circuit consisting of a capacitor and a shunt resistor, both parallel to the wire coil indicating the self inductance that is the tendency of an electrical conductor to oppose a change in the electric current flowing through it.

The quality factor of the resonance reflects the damping coefficient, which is given by $Q:=R\big(\frac{C}{L_J}\big)^{\frac{1}{2}}=R\big(\frac{2eI_cC}{\hbar}\big)^{\frac{1}{2}}$. When $Q$ exceeds
$\frac{1}{2}$, the plasma resonance is underdamped. Conversely, if $Q$ falls below $\frac{1}{2}$, the plasma resonance is overdamped.
The plasma frequency $\omega_{p}:=\frac{Q}{RC}=(CL_J)^{-\frac{1}{2}}=\big(\frac{2eI_c}{C\hbar}\big)^{\frac{1}{2}}$ of the junction  is determined by
the resonant frequency of RLC circuit. After introducing the dimensionless scaling parameter $\omega_{p}$, we rescale the time by it as
$t=\omega_{p}\tau$. Thus the system  \eqref{Newtonseq} becomes
\begin{equation*}
\frac{d^{2}\phi}{dt^{2}}=-\beta_{J}\frac{d\phi}{dt}-\sin(\phi)+\kappa,
\end{equation*}
where $\beta_{J}:=\frac{1}{Q}=G\big(\frac{\hbar}{2eI_cC}\big)^{\frac{1}{2}}$ is the dissipation coefficient, which stands for a parameter inversely related to the Josephson plasma frequency $\omega_{p}$. Additionally, $\kappa:=I_{S}/I_c$ represents the amplitude of the current bias normalized to the critical current.

\begin{figure}
\begin{center}
\begin{minipage}{2in}
\leftline{(a)}
\includegraphics[width=2in]{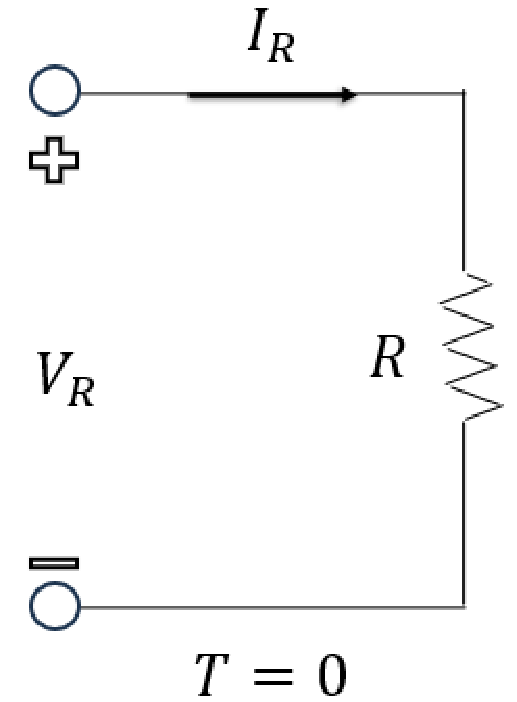}
\end{minipage}
\hfill
  \begin{minipage}{3.5in}
\leftline{(b)}
\includegraphics[width=3.5in]{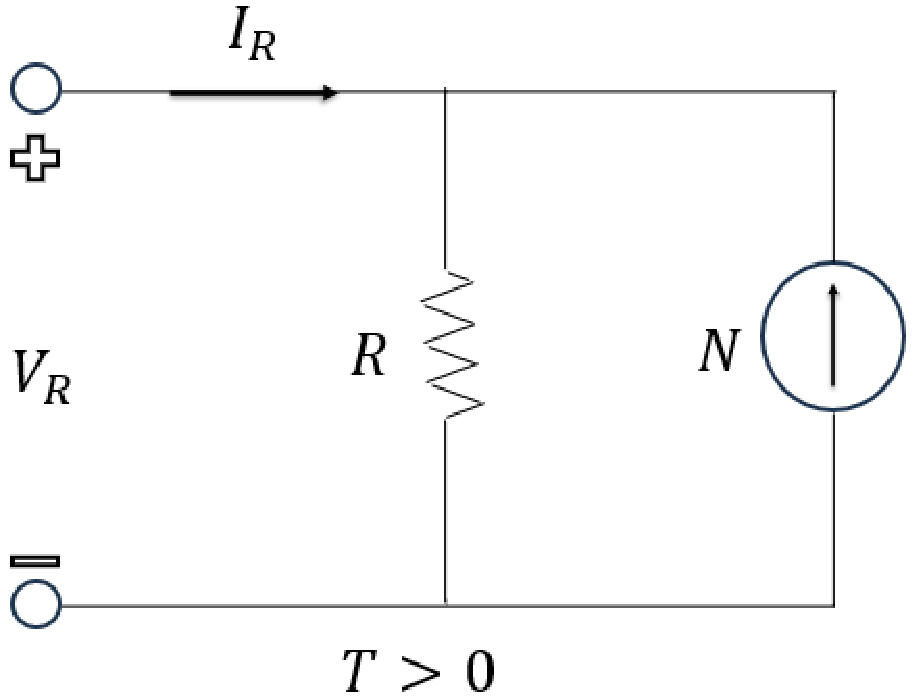}
\end{minipage}
\caption{(a) There exists a resistor $R$ in the absence of thermal noise when the temperature $T$ of the circuit is set to zero; (b) At any non-zero temperature, the resistor is replaced by a parallel combination of resistor $R$ and current noise source $N$, as dictated by the fluctuation-dissipation theorem.}\label{FD}
\end{center}

\end{figure}

In Figure \ref{FD}(a), there is no  thermal noise if the temperature $T$ vanishes from the circuit. A resistor $R$ exists in the absence of thermal noise at zero temperature. While Figure \ref{FD}(b) describes that
at any finite temperature the resistor is replaced by the parallel combination of a shunt resistor and a thermal noise $N(\tau)$ based on the fluctuation-dissipation theorem.
The current source $N(\tau)$ related to the damping represents a white Gaussian noise process with zero mean and autocorrelation function
\begin{equation}\label{NN}
\langle N(\tau_1)(\tau_2)\rangle=\frac{2k_{B}T}{R}\delta(\tau_1-\tau_2),
\end{equation}
where the Dirac delta function $\delta$ in time indicates that there is no correlation between impacts in
any distinct time intervals $d\tau_1$ and $d\tau_2$. The thermal energy $k_{B}T$ is the thermostat of Boltzmann constant $k_{B}$ and temperature $T$.
In terms of dimensionless quantities, equation \eqref{NN} is equivalent to the dimensionless noise with a zero mean value and a second moment given by
\begin{equation*}
\langle\zeta(t_1)\zeta(t_2)\rangle=2\beta_J\Gamma\delta(t_1-t_2),
\end{equation*}
where $\zeta=N/I_c$ is the normalized noise current, and
\begin{equation*}
\Gamma:=\frac{k_BT}{E_J}=\frac{\text{the thermal energy}}{\text{the Josephson coupling energy}}
\end{equation*}
corresponds to the dimensionless temperature. With this noise source added to the circuit, the equation of motion becomes
\begin{equation}\label{fine}
\frac{d^{2}\phi}{dt^{2}}+\beta_{J}\frac{d\phi}{dt}+\sin(\phi)=\kappa+\zeta(t).
\end{equation}

\section{Stochastic dynamics}\label{DRSTJ}
Explicitly, we consider the resistively shunted superconducting tunnel junction driven by the thermal noise according to the stochastic model

\begin{equation}\label{STJ}
\ddot{x}+\beta_{J}\dot{x}+\sin x=\kappa+\sqrt{2D}\eta(t),
\end{equation}
where $x$ represents a phase difference influenced by thermal fluctuation, $\kappa=I_{S}/I_{c}$ is related to the total current across the
junction, $\beta_{J}=G\big(\frac{\hbar}{2eI_cC}\big)^{\frac{1}{2}}=\frac{1}{Q}$ stands for a parameter inversely related to the
Josephson plasma frequency $\omega_{p}=(CL_J)^{-\frac{1}{2}}=\big(\frac{2eI_c}{C\hbar}\big)^{\frac{1}{2}}$, and
\begin{equation*}
 \frac{dx}{d\tau}=\omega_{p} \frac{dx}{dt}.
\end{equation*}
The parameters can be changed during the simulation.
The overdot denotes differentiation with respect to time $t$.
The interaction with thermostat  is modelled by $\delta$-correlated, Gaussian white
noise $\eta(t)$ of vanishing mean and unit intensity, i.e.,
\begin{equation}\label{thermostat}
\langle\eta(t)\rangle=0,\quad\quad\langle\eta(t)\eta(s)\rangle=\delta(t-s).
\end{equation}
The processes $\eta(t)$ and $\frac{1}{\sqrt{2D}}\zeta(t)$ have the same distribution, where the noise intensity $D:=\beta_J\Gamma$ is a measure of the strength of the fluctuating force. So the solution of stochastic system \eqref{STJ} coincides with that
of system \eqref{fine} in distribution.

The resistively shunted superconducting tunnel junction model in the absence
of noise is
\begin{equation}\label{JJ}
\ddot{x}+\beta_{J}\dot{x}+\sin x=\kappa.
\end{equation}
We define a new variable $v:=\dot{x}$. Then the second-order differential equation \eqref{JJ} can be transformed into a planar system of the form
\begin{equation}\label{planar}
\left\{
  \begin{array}{ll}
    \dot{x}=v,  \\
    \dot{v}=\kappa-\beta_{J}v-\sin x,
  \end{array}
\right.
\end{equation}
where $\xi:=\beta_{J}v$ is proportional to the voltage across the junction.

 When $\kappa=\beta_{J}=0$, the dynamical system \eqref{planar} becomes a Hamiltonian system having one degree of freedom given by
\begin{equation*}
H(x,v)=\frac{v^2}{2}-\cos x,\quad\text{(kinetic energy+potential energy)}
\end{equation*}
and Hamilton's equations are as follows:
\begin{equation*}
\left\{
  \begin{array}{ll}
  \dot{x}=v,\\
\dot{v}=-\sin x.
  \end{array}
\right.
\end{equation*}
The equilibria of the above planar system  are points where
$\frac{\partial H}{\partial x}=\sin x=0$ and $\frac{\partial H}{\partial v}=v=0$.
Thus the fixed points occur at $(n\pi,0)$ in the $(x,v)$ plane, where $n$ is an
integer. The Jacobian matrix is
\begin{equation*}
J=\left(
            \begin{array}{cc}
              0 & 1 \\
              -\cos x & 0 \\
            \end{array}
          \right).
\end{equation*}
 For $m\in\mathbb{Z}$, the equilibria at $\big((2m+1)\pi,0\big)$ are saddle points with eigenvalues $\lambda=\pm1$ of $J$ when $n=2m+1$, while the equilibria at $(2m\pi,0)$ are topological centers with eigenvalues $\lambda=\pm i$ of $J$ when $n=2m$.
 The fixed points are hyperbolic if $n$
is odd and nonhyperbolic if $n$ is even. Note that the Hartman-Grobman theorem requires that the equilibrium must be hyperbolic. Therefore, Hartman-Grobman theorem cannot
be applied when $n$ is even.

\begin{figure}
\begin{center}
\begin{minipage}{2.6in}
\leftline{(a)}
\includegraphics[width=2.6in]{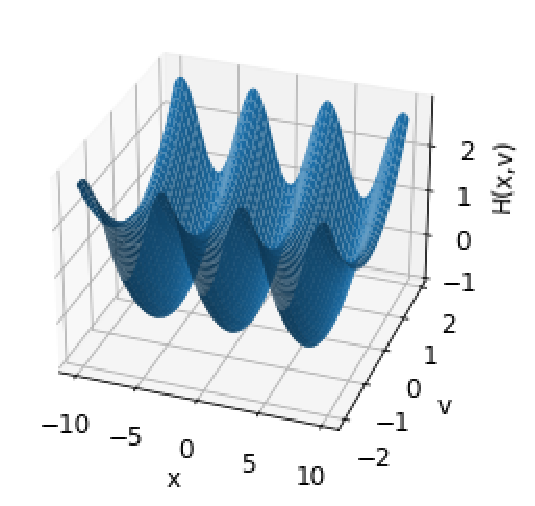}
\end{minipage}
\hfill
  \begin{minipage}{3.7in}
\leftline{(b)}
\includegraphics[width=3.7in]{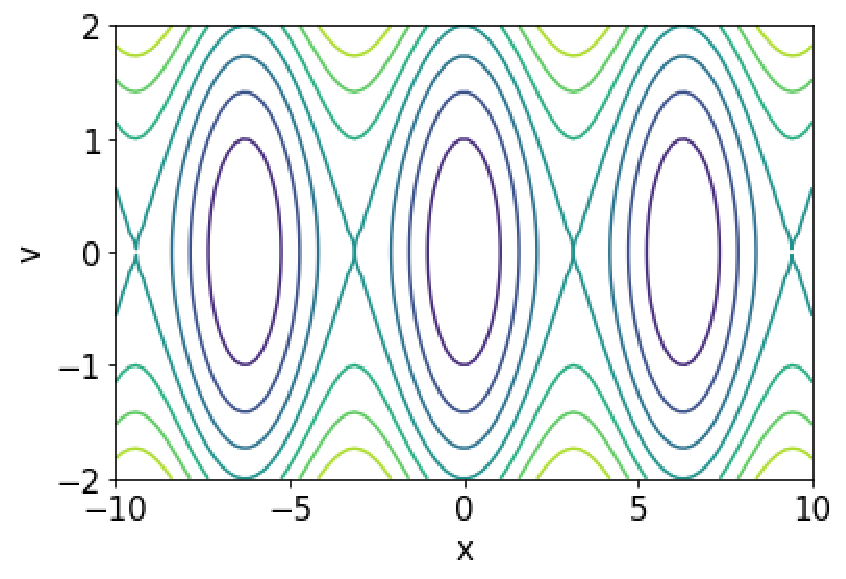}
\end{minipage}
\caption{(a) The energy
surface $H(x,v)$; (b) A phase portrait of a Hamiltonian system is plotted in the $(x,v)$ plane, where the vertical axis represents
$v$, and the horizontal axis shows $x$.}\label{HC}
\end{center}
\end{figure}

Since the Hamiltonian energy is conserved, the motion is along contours of the energy $H$
in the phase space $(x,v)$.  Figure \ref{HC}(a) shows the energy diagram of a superconducting
tunnel junction. Solution curves are sketched in
Figure \ref{HC}(b) for which the axes are the displacement $x$ and angular
velocity $v$. The closed curves in Figure \ref{HC}(b) characterize local minima on the surface $H(x,v)$ as in Figure \ref{HC}(a), and
the unstable fixed points correspond to local maxima on the same surface. The rotating curves surrounding the topological centers $(2m\pi,0)$ represent bounded and periodic oscillations.
We find that the heteroclinic connections at the saddle
points $\big((2m+1)\pi,0\big)$ form the separatrix.
The wavy lines for large velocities describe librating motions.% in which (\textcolor[rgb]{1.00,0.00,0.00}{the pendulum spins around its pivotal point})

\begin{figure}
\begin{center}
\begin{minipage}{3.1in}
\leftline{(a)}
\includegraphics[width=3.1in]{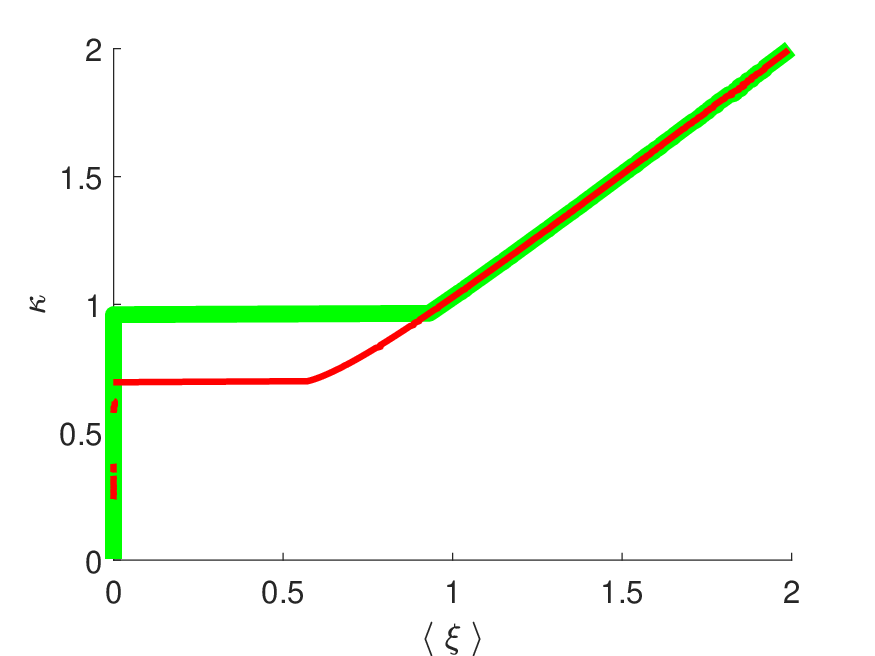}
\end{minipage}
\hfill
  \begin{minipage}{3.3in}
\leftline{(b)}
\includegraphics[width=3.3in]{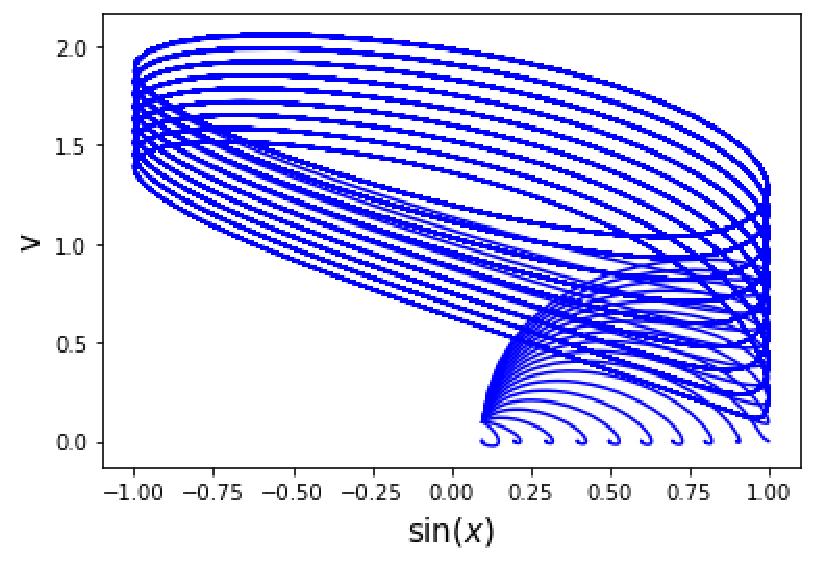}
\end{minipage}
\caption{(a) A typical I-V characteristic curve is simulated for a superconducting tunnel Josephson junction.
It exhibits a clockwise hysteresis. The Cooper pair tunneling
current appears when $\langle\xi\rangle=0$, while the
quasiparticle tunneling current is
detected when $\langle\xi\rangle>2\Delta/e$. Note that $\kappa=I_S/I_c$, where $I_S$ represents the current through the Josephson junction, and $I_c$ is a parameter of the junction known as the critical current; (b) An animation of the bifurcating limit cycle.}\label{JB}
\end{center}
\end{figure}

For nonzero $\kappa$ and $\beta_J$, interesting dynamics are observed. We
investigate some of these via the sketch of typical current-voltage (I-V) characteristic curve of a superconducting tunnel junction, as depicted in Figure \ref{JB}(a) for $\beta_J=0.6$. There is a clockwise hysteresis. The vertical axis is $\kappa=I_S/I_c$, where $I_c$ is the critical current of the junction. The horizontal axis corresponds to the normalized direct current
voltage $\langle\xi\rangle=\beta_{J}\langle v\rangle$, where $\langle v\rangle$ is the average of the maximum and minimum
values of $v$ in the long-time region with  $t\in[0,200]$.

The red curve reflects the current for decreasing
voltage. As the voltage $\langle\xi\rangle$  is
decreased down to the gap voltage, the relation $\kappa=\langle\xi\rangle$ still holds until a critical state at which $\kappa_c\approx0.6965$ and
$\langle\xi\rangle=2\Delta/e$, where $\Delta$ is an energy gap of the superconductor and $e$ is the electron charge.
Since the bias voltage $e\langle\xi\rangle$ exceeds
twice the quantized value of superconducting energy gap $\Delta$, the
quasiparticle current arises, i.e., $\kappa>\kappa_c$. The red bar at $\kappa=\kappa_c$ represents the  direct current Josephson effect. There is a bifurcation from oscillatory behavior back to the zero-voltage point $\langle\xi\rangle=0$. A direct current Josephson supercurrent flows under $\langle\xi\rangle=0$.

The green curve depicts the current for increasing voltage. When $\langle\xi\rangle=0$, Josephson current $I_S$ flows up to a maximum threshold value $I_c$, i.e., $\kappa<1$. If $0<\langle\xi\rangle<1$, then $\kappa=1$, i.e., $I_S=I_c$. As
$\langle\xi\rangle$ is increased further, the identity $\kappa=\langle\xi\rangle$ is valid when
the current bias $I_S$ exceeds the critical current $I_c$, i.e., $\kappa>1$. There is a bifurcation to an oscillating tunneling current.

The fixed points of the dynamical system \eqref{planar} of first-order autonomous differential equations are found by solving the equations $\dot{x}=\dot{v}=0$.
Set $v=0$. Then $\dot{x}=0$. It follows that $\dot{v}=0$ if and only if $\sin x=\kappa$. If the current bias is less than the critical current, i.e., $\kappa<1$,
two equilibria  are calculated as $(x,v)=(x_{e},0)$ and $(\pi-x_{e},0)$, where $x_{e}=\sin^{-1}\kappa$, $0<x_{e}<\pi/2$. The first solution is stable and represents the superconducting state, whereas the second one is an unstable saddle point.
Taking initial values of $\sin x_0=0.1$ and $v_0=0.1$, we plot an animation of a limit cycle bifurcation using $\beta_J=1.2$ as the parameter $\kappa$ increases from $\kappa=0.1$ to $\kappa=2$ by step 0.1, as demonstrated in Figure \ref{JB}(b). The horizontal axis measures $\sin x$; the vertical axis represents voltage.

\begin{figure}[h]
\begin{center}
  \begin{minipage}{2.13in}
\leftline{(a)}
\includegraphics[width=2.13in]{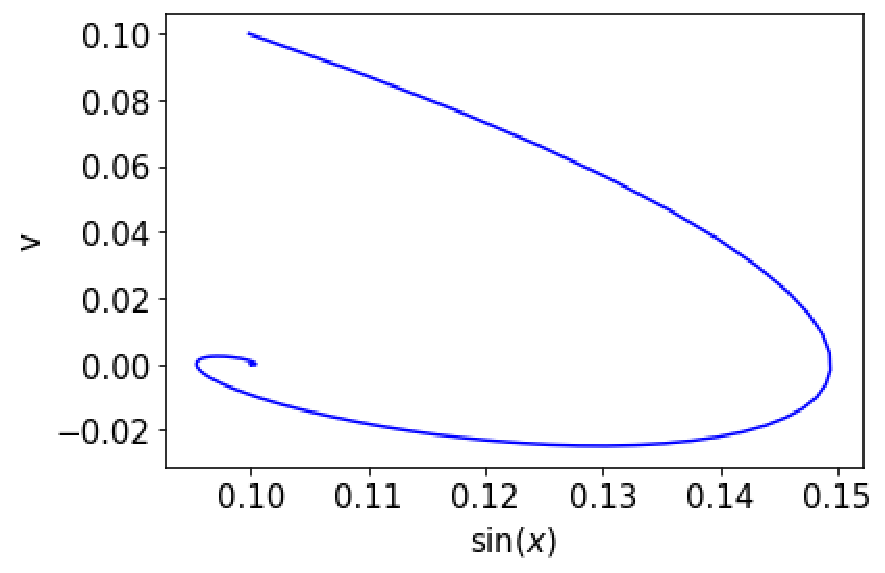}
\end{minipage}
\hfill
\begin{minipage}{2.13in}
\leftline{(b)}
\includegraphics[width=2.13in]{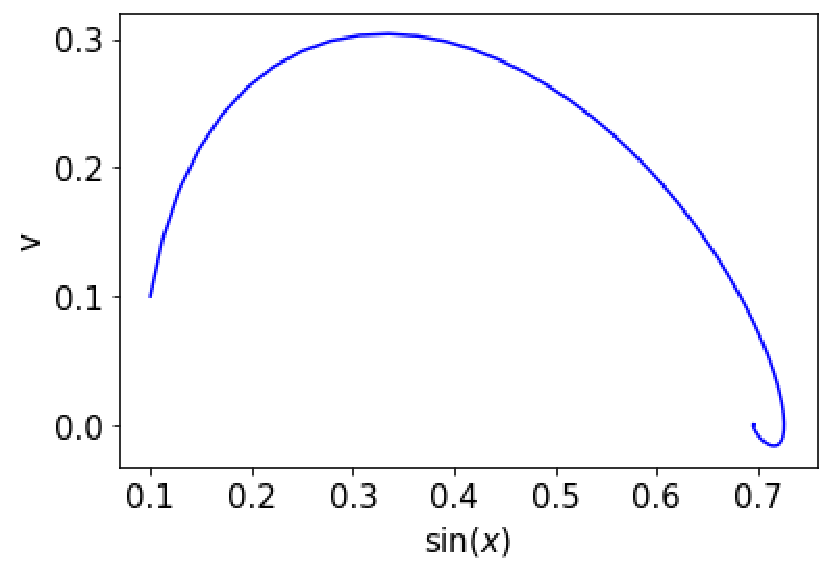}
\end{minipage}
\hfill
  \begin{minipage}{2.13in}
\leftline{(c)}
\includegraphics[width=2.13in]{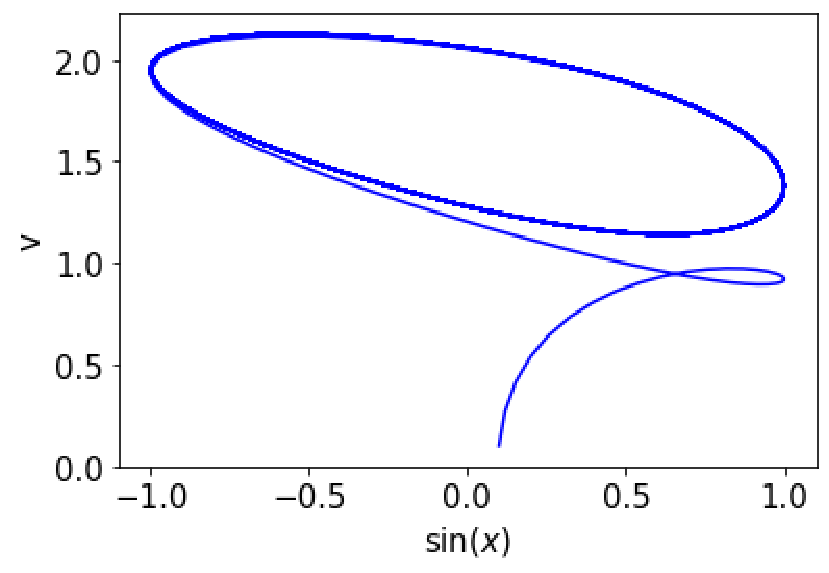}
\end{minipage}
\caption{ Bifurcation of a planar limit cycle in a resistively shunted Josephson junction with $\beta_J=1.2$ : (a) $\kappa=0.1$; (b) $\kappa=0.6965$; (c) $\kappa=2$. }\label{JA}
\end{center}
\end{figure}

%(0, tmax=100, 0.1)
 For fixed $\beta_J=1.2$, we take three snapshots of the animation for different values of $\kappa$ to describe the bifurcations that occur in the model \eqref{planar} since the limit cycle appears. When $\kappa=0.1$, the path goes from the starting point $(\sin x_0,v_0)=(0.1,0.1)$ to the ending point  $(\sin x,v)=(0.1,0)$ , as illustrated in Figure \ref{JA}(a). As can be seen from Figure \ref{JA}(b), the blue curve moves from the initial position $(\sin x_0,v_0)=(0.1,0.1)$ to the final position $(\sin x,v)=(0.6965,0)$ in the case of $\kappa=0.6965$.  Figure \ref{JA}(c) shows a limit cycle in a resistively shunted Josephson junction when $\kappa=2$. The trajectory starting from the point $(\sin x_0,v_0)=(0.1,0.1)$ reaches this limit cycle.

%The graph of the function
%$y=2-\sin x$ has no roots.  If $\kappa=2$ and $\beta_J=1.2$, the system \eqref{planar} has no fixed points. Therefore, there exists a unique limit cycle by the corollary to the
%Poincar\'e-Bendixson theorem \cite{P}.

\begin{figure}[h]
\begin{center}
  \begin{minipage}{2.13in}
\leftline{(a)}
\includegraphics[width=2.13in]{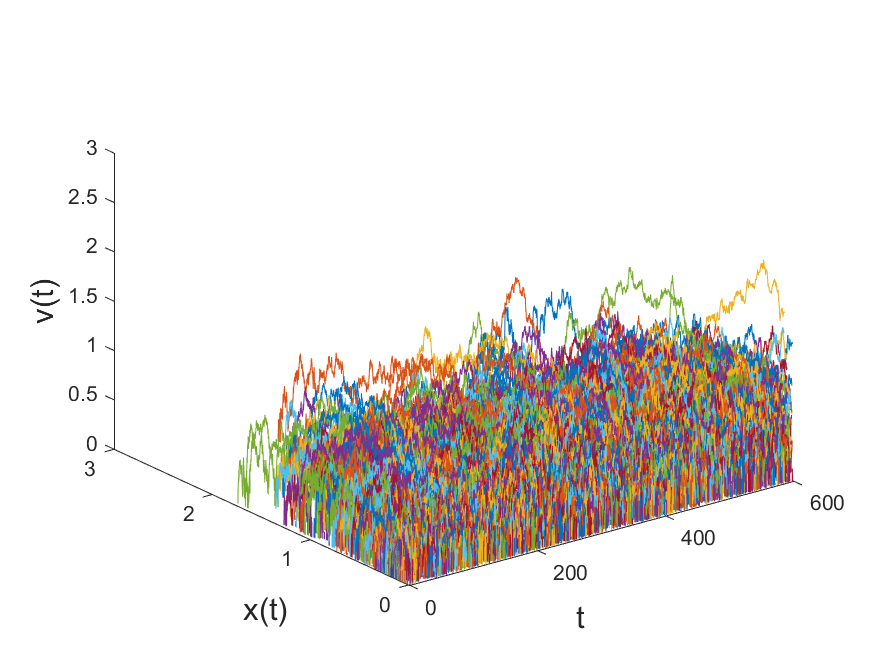}
\end{minipage}
\hfill
\begin{minipage}{2.13in}
\leftline{(b)}
\includegraphics[width=2.13in]{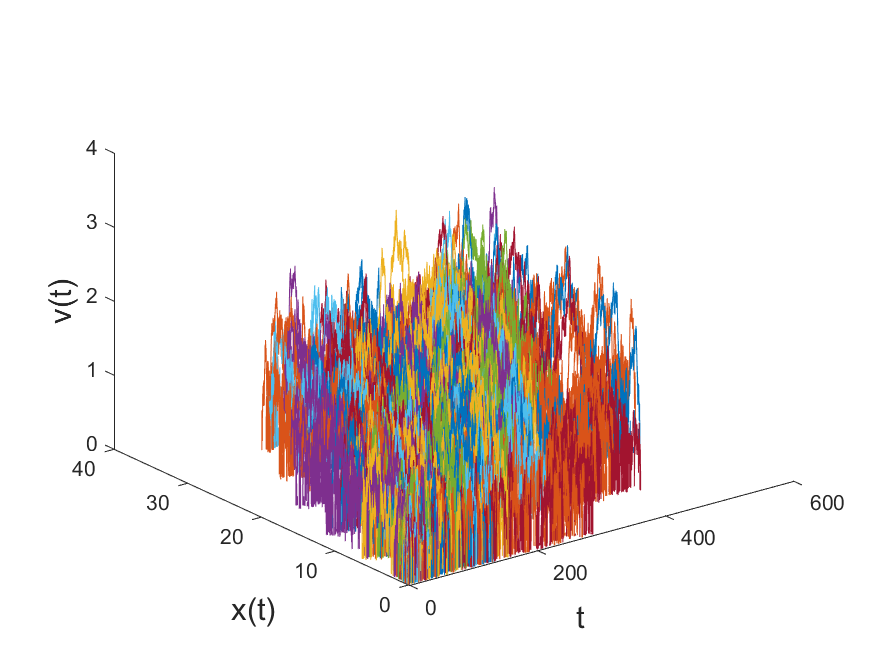}
\end{minipage}
\hfill
  \begin{minipage}{2.13in}
\leftline{(c)}
\includegraphics[width=2.13in]{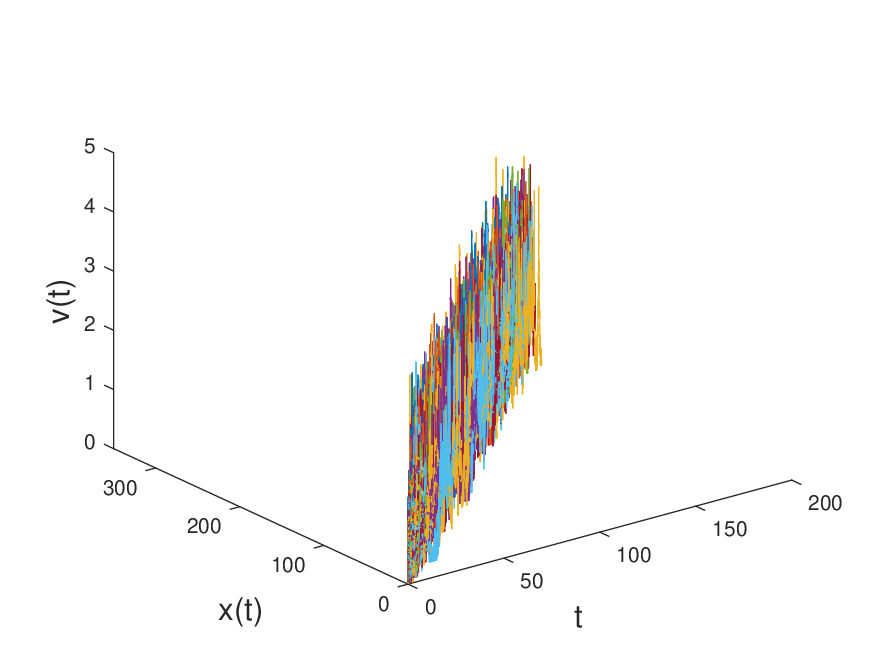}
\end{minipage}
\caption{The evolution of the variables $x(t)$ and $v(t)$ with respect to time characterized by system  \eqref{STJ} for $\beta_J=1.2$:
 (a) When $\kappa=0.1$ and $t$  ranges from 0 to 600, the variable $x$ varies from 0 to 3, whereas in the stochastic sample trajectories, the variable $v$ randomly decreases from 2 to 0; (b) When $\kappa=0.6965$ and within the time range of 0 to 400, the value of $x(t)$ changes from 0 to 40, while the value of $v(t)$ fluctuates between 0 and 3. The resulting trajectories of this motion take on the shape of a fluctuating square; (c) When $\kappa=2$ and $t$ varies from 0 to 200, the time $x(t)$ evolves from 0 to 350, while concurrently, the function $v(t)$ oscillates rapidly within the range of 0 to 4. The functions $x(t)$ and $v(t)$ together form a wall exhibiting a random diagonal pattern during this time interval.}\label{xvt}
\end{center}
\end{figure}
\begin{figure}[h]
\begin{center}
  \begin{minipage}{2.13in}
\leftline{(a)}
\includegraphics[width=2.13in]{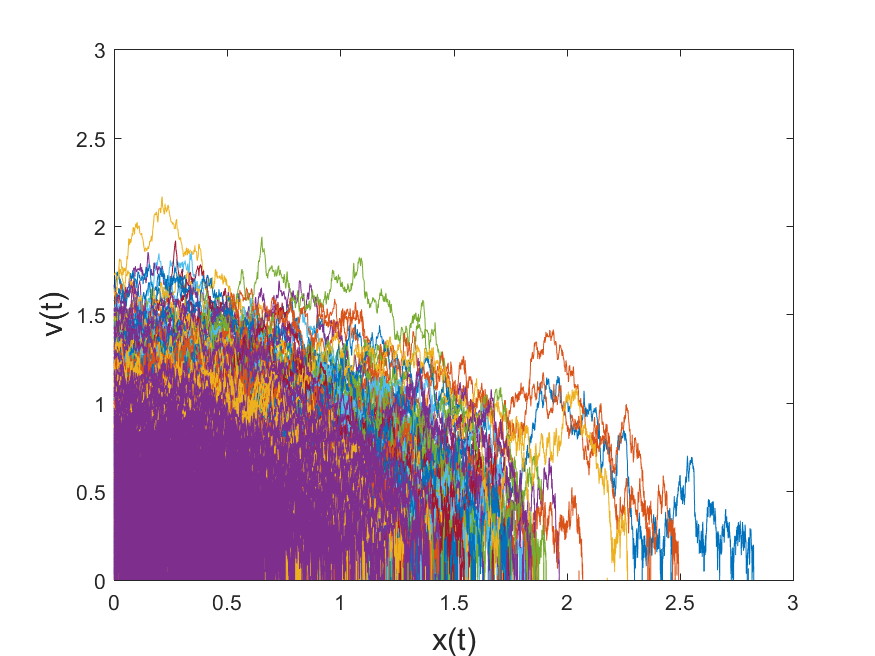}
\end{minipage}
\hfill
\begin{minipage}{2.13in}
\leftline{(b)}
\includegraphics[width=2.13in]{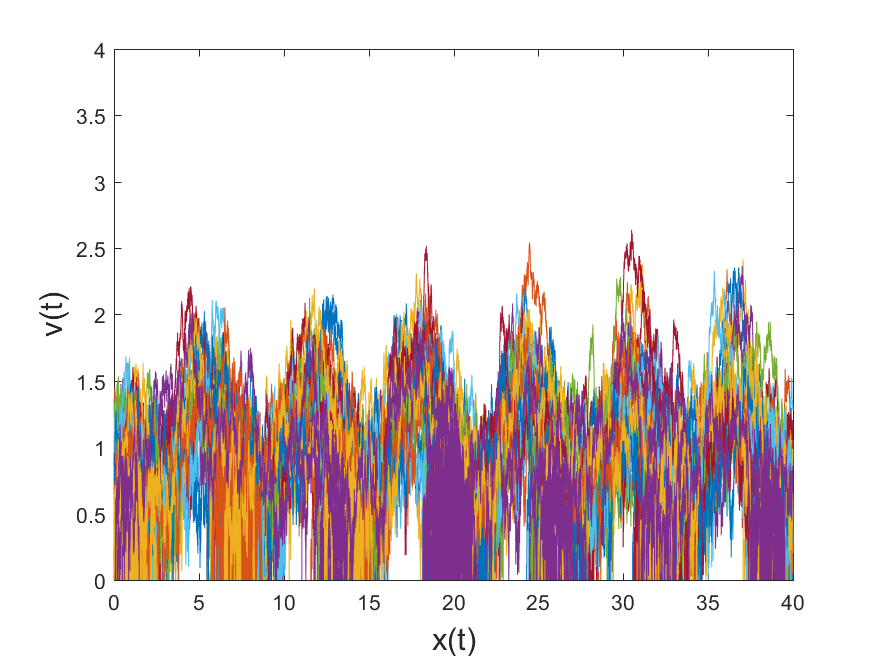}
\end{minipage}
\hfill
  \begin{minipage}{2.13in}
\leftline{(c)}
\includegraphics[width=2.13in]{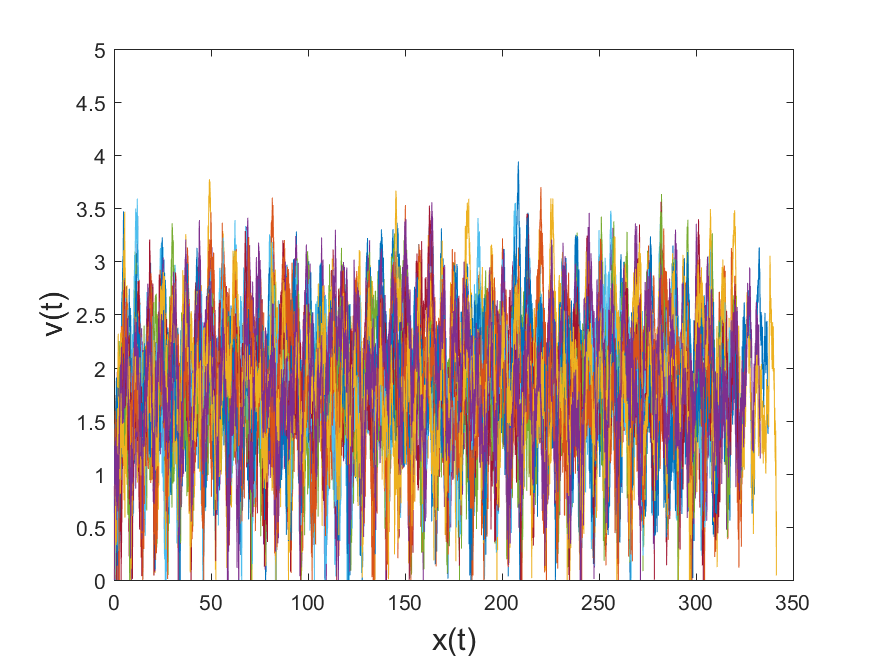}
\end{minipage}
\caption{The interaction between the variables $x(t)$ and $v(t)$ described by system \eqref{STJ} with $\beta_J=1.2$: (a) When $\kappa=0.1$, for most trajectories, $v(t)$ radiates from 2 to 0 when $x(t)$ reaches 2. However, for only a small fraction of trajectories, $v(t)$ radiates to 0 when $x(t)$ reaches 2.8;  (b) When $\kappa=0.6965$, $x(t)$ varies from 0 to 40, and $v(t)$ fluctuates within the range of 0 to 3. The resulting trajectories oscillate within a stochastic wave band; (c) When $\kappa=2$, $x(t)$ evolves from 0 to 350, and $v(t)$ oscillates rapidly within the range of 0 to 4.}\label{xv}
\end{center}
\end{figure}
\begin{figure}[h]
\begin{center}
  \begin{minipage}{2.13in}
\leftline{(a)}
\includegraphics[width=2.13in]{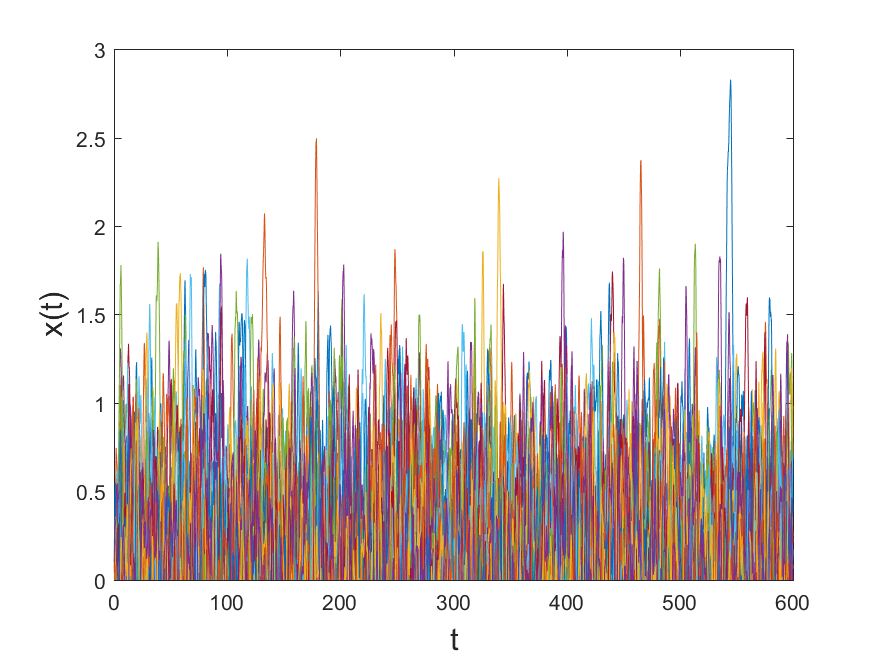}
\end{minipage}
\hfill
\begin{minipage}{2.13in}
\leftline{(b)}
\includegraphics[width=2.13in]{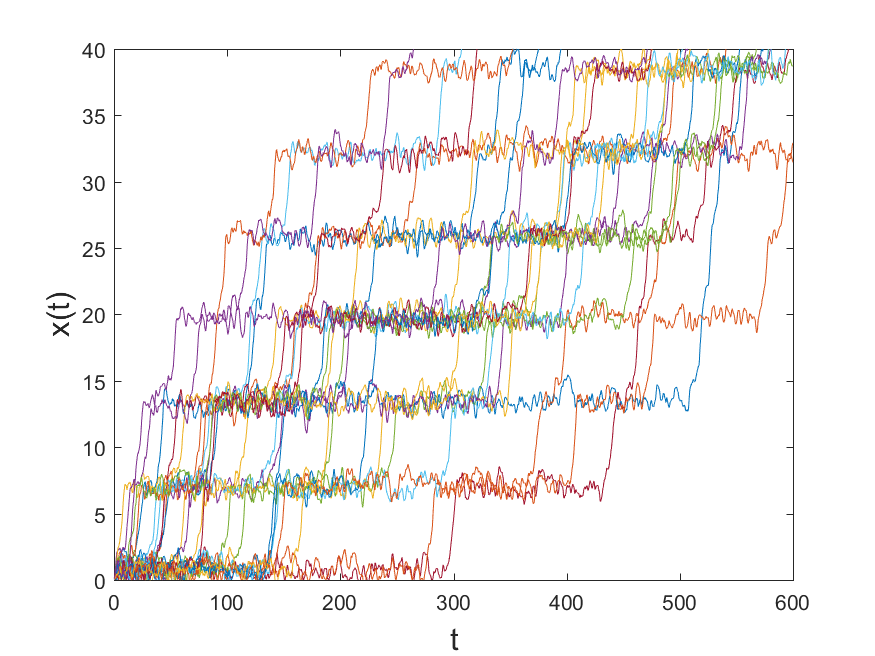}
\end{minipage}
\hfill
  \begin{minipage}{2.13in}
\leftline{(c)}
\includegraphics[width=2.13in]{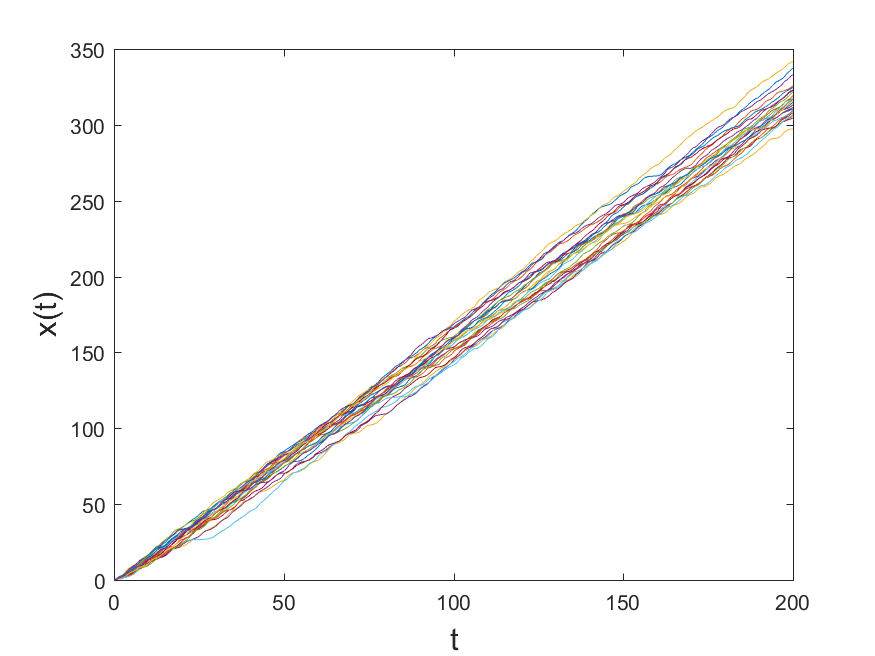}
\end{minipage}
\caption{In the stochastic superconducting system governed by Eq. \eqref{STJ} with a parameter $\beta_J=1.2$: (a) When $\kappa=0.1$, the trajectories of the displacement $x(t)$ within the range [0,3] are multiplexed over the time range $t\in[600]$, generating a random pattern resembling grass; (b) When $\kappa=0.6965$, the trajectories of the displacement $x(t)\in[0,40]$ form a tilted washboard pattern over the time range $t\in[600]$. Each of the seven steps within this pattern displays random amplitudes; (c)
When $\kappa=2$, the trajectories of the displacement $x(t)\in[0,350]$  resemble a randomly fluctuating diagonal beam within the time range $t\in[200]$.
}\label{xt}
\end{center}
\end{figure}

After specifying the equivalent noise index $\sqrt{2D}=0.5$ and the parameter value $\beta_J=1.2$ in system \eqref{STJ}, we reasonably use the Monte Carlo method to perform detailed numerical simulations of dynamical orbits. Based on Figures \ref{xvt}-\ref{xt}, we classify and identify transmission patterns in random media. In Figure \ref{xvt} (a), considering the case where parameter $\kappa=0.1$, the random disorder of the three-dimensional orbital evolution between time, displacement, and angular velocity is significantly strong. Figure \ref{xvt} (b) shows a three-dimensional random block distribution of time, displacement, and angular velocity in a completely new form near the critical parameter $\kappa=0.6965$. When $\kappa=2$, the three-dimensional orbits between time, displacement, and angular velocity in Figure \ref{xvt} (c) are arranged along a diagonal random wall.

Figure \ref{xv} (a) reflects the radioactive random orbit between displacement and angular velocity at $\kappa=0.1$. The interaction orbit between displacement and angular velocity simulated in Figure \ref{xv} (b) at $\kappa=0.6965$ exhibits a periodic random band with wavelength, amplitude, phase, polarization, etc. In Figure \ref{xv} (c), the image between displacement and angular velocity in the stochastic superconducting system \eqref{STJ} is like a row of needle tips at $\kappa=2$.

 For an intuitive numerical simulation we plot the evolutionary trajectory of the superconducting phase perturbed by thermal noise  as a function of time. Unexpectedly, the displacement trajectory is multiplexed over time, resembling a mixing module of random grass, see Figure \ref{xt} (a) at $\kappa=0.1$. If the parameter grows to $\kappa=0.6965$, the displacement trajectory forms a random laundry plate nonlinear microwave quantum network with wavelength switching, as suggested by Figure \ref{xt} (b). Figure \ref{xt} (c) depicts the trajectory of the displacement in a stochastic superconducting system \eqref{STJ} with parameter $\kappa=2$ changing randomly along a diagonal line, resulting in a large beam coverage range.

\begin{figure}[h]
\begin{center}
  \begin{minipage}{2.13in}
\leftline{(a)}
\includegraphics[width=2.13in]{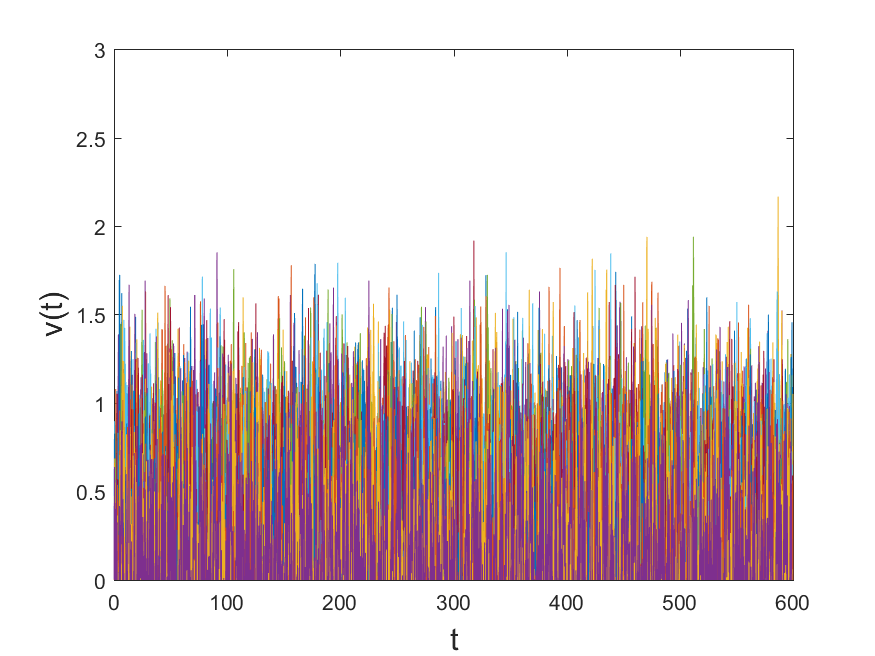}
\end{minipage}
\hfill
\begin{minipage}{2.13in}
\leftline{(b)}
\includegraphics[width=2.13in]{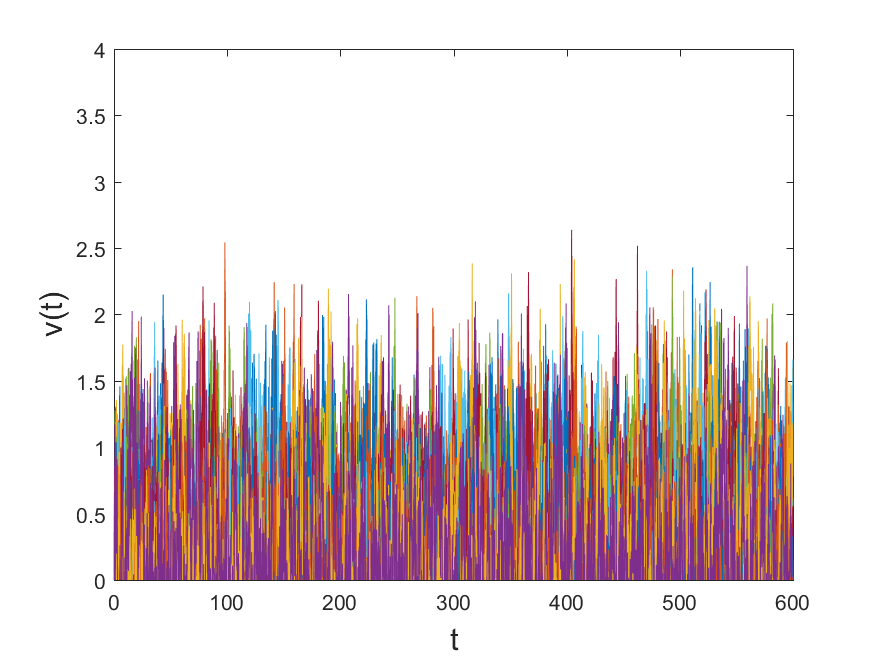}
\end{minipage}
\hfill
  \begin{minipage}{2.13in}
\leftline{(c)}
\includegraphics[width=2.13in]{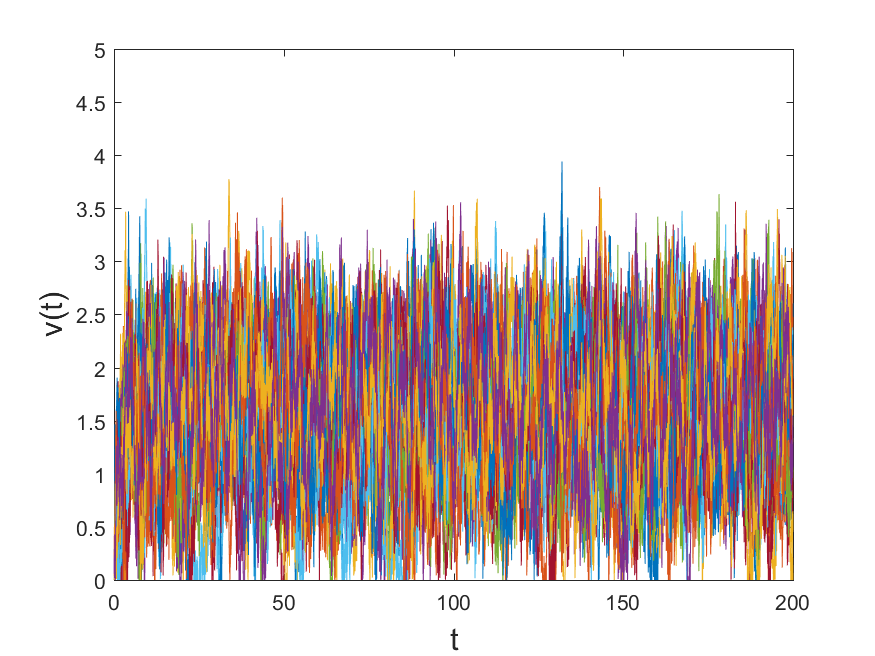}
\end{minipage}
\caption{In the stochastic superconducting system \eqref{STJ} with a specific parameter value of $\beta_J=1.2$: (a) When $\kappa=0.1$, the evolution of voltage  $v(t)$ can reach a height of  2 within the time interval  $t\in[0,600]$. This evolution resembles a row of blades of grass, with a relatively dense concentration between 0 and 1, and a relatively sparse distribution between 1 and 2; (b) When $\kappa=0.6965$, the voltage evolution $v(t)$ can attain a height of 2.5 over the time range $t\in[0,600]$, exhibiting a resemblance to a row of grassy blades. The concentration is denser between 0 and 1.5, and becomes relatively sparse between 1.5 and 2.5; (c) When $\kappa=2$, the voltage evolution $v(t)$ can attain a height of 4 within the time range $t\in[0,200]$, displaying a resemblance to a grassy distribution. During this period, the density of the voltage evolution is highest between 0.5 and 3, and becomes relatively sparse both below 0.5 and above 3, up to the height of 4.}\label{vt}
\end{center}
\end{figure}

For the parameters $\kappa=0.1$, 0.6965 and 2, the voltage across the Josephson junction under external excitation presents different characteristics with time, we can observe the following phenomena in Figure \ref{vt}. As exhibited in Figure \ref{vt}(a), when $\kappa=0.1$, the voltage evolution height can reach 2 over time, which is similar to a row of blades of grass, relatively dense between 0 and 1, and relatively sparse between 1 and 2. This indicates that the response of the Josephson junction is more concentrated in a smaller voltage range at lower parameters, and becomes relatively sparse as the voltage increases.
When the parameter  $\kappa$ increases to 0.6965, the voltage evolution height over time reaches 2.5, which still is like a row of grassy blades. It is concentrated between 0 and 1.5, while relatively sparse between 1.5 and 2.5, as rendered in Figure \ref{vt}(b). But The dense area of response is slightly larger than that in the previous case. This infers that the response of the Josephson junction becomes dense over a larger voltage range as the parameter $\kappa$ increases.
 When $\kappa=2$, the voltage evolution height over time can reach 4,  which still presents a grass-like distribution.  At this point, the response is dense between 0.5 and 3, while relatively sparse between 0 and 0.5 and between 3 and 4, as pictured in Figure \ref{vt}(c). Compared with the first two parameter values, when the parameter $\kappa$ is further increased to 2, the dense region of the response expands to a larger voltage range, but it is concentrated in the middle voltage range.
These observations show that the voltage evolution of the Josephson junction under external excitation exhibits different distribution characteristics over time.
  These characteristics change as the parameter $\kappa$ increases.
  At lower parameters, the response is concentrated in a smaller voltage range. However, with the parameters increasing, the response gradually expands to a larger voltage range while still maintaining a certain concentration area. The observed distribution may be influenced by both the intrinsic mechanism of the Josephson junction and the properties of external incentives.

\section{Fokker-Planck equation}\label{FPE}

The mapping defined by equation \eqref{STJ} can also be thought of as two-dimensional. Rewrite the second-order SDE \eqref{STJ} into two-dimensional first-order SDEs  \begin{equation*}
\left\{
\begin{array}{ll}
dx=vdt,  \\
dv=(\kappa-\beta_{J}v-\sin x)dt+\sqrt{2D}d\eta(t), \end{array}
\right.
\end{equation*}
where the thermal noise voltage source is Brownian motion. The state of the above system is specified by two variables $x$  and $v=dx/dt$.
At finite temperatures, these variables evolve according to a two-dimensional continuous Markov process, influenced by Brownian noise.
The equation of motion in abbreviated form is
\begin{equation}\label{abbreviated}
\frac{d\textbf{u}}{dt}=\textbf{F}(\textbf{u})+\Sigma\Theta(t),
\end{equation}
where the quantities are
\begin{equation*}
\textbf{u}=\left(
             \begin{array}{c}
               u \\
               v \\
             \end{array}
           \right),\quad
\textbf{F}=\left(
             \begin{array}{c}
               v \\
               \kappa-\beta_{J}v-\sin x\\
             \end{array}
           \right),\quad \Sigma=\left(\begin{array}{cc}
                             0 & 0 \\
                             0 & \sqrt{2D}
                           \end{array}\right),\quad \Theta(t)=\left(
                                                      \begin{array}{c}
                                                        0 \\
                                                        \eta(t) \\
                                                      \end{array}
                                                    \right).
\end{equation*}

Within the framework of statistical mechanics, the time evolution is characterized predominantly by the phase space distribution function $p(\textbf{u},t)$ that determines the average properties of the system being in a specific state $\textbf{u}$ at a particular time $t$. The probability of finding the system state within the region $d\textbf{u}$ surrounding the point $\textbf{u}$ at time $t$ is given by $p(\textbf{u},t)d\textbf{u}$. Rather than seeking a general solution to the equation \eqref{abbreviated}, we are interested in the probability density $p(\textbf{u},t)$ for the value of $\textbf{u}$ at time $t$. Furthermore, what we really want is the average of this probability distribution over the noise. One way to find the noise average is to recognize that $p(\textbf{u},t)$ is a conserved quantity, ensuring that the total probability of describing the system somewhere is unity at all times. Mathematically, this is expressed as an integral over all possible states $\textbf{u}$ of the phase space distribution function $p(\textbf{u},t)d\textbf{u}$, which must add up to 1 at any given time. It means that
\begin{equation*}
\int p(\textbf{u},t)d\textbf{u}=1,\quad \forall\,\,t.
\end{equation*}

 The above conservation law is a fundamental principle in probability theory, which is crucial for maintaining that the description of the system's behavior is consistent and complete, making it a valuable tool in statistical mechanics and related fields. We expect that the time derivative of the conserved density $p(\textbf{u},t)$ is balanced by the divergence of a flux, which is the product of a velocity and that density. This is how the Liouville equation
\begin{equation*}
\frac{\partial p}{\partial t}+\frac{\partial}{\partial \textbf{u}}\Big(\frac{d\textbf{u}}{dt}p\Big)=0,
\end{equation*}
is derived in statistical mechanics. Here, the space coordinate is $\textbf{u}$, the density at $\textbf{u}$ is $p(\textbf{u},t)$, and the velocity at $\textbf{u}$ is $d\textbf{u}/dt$. When the time derivative $d\textbf{u}/dt$ is replaced with the right-hand side of \eqref{abbreviated},
we get
\begin{equation*}
\frac{\partial p(\textbf{u},t)}{\partial t}=-\frac{\partial}{\partial\textbf{u}}\big(\textbf{F}(\textbf{u})p(\textbf{u},t)+\Sigma\Theta(t)p(\textbf{u},t)\big).
\end{equation*}
Therefore, the Fokker-Planck equation describing the probability density $p(x,v,t)$ for the phase-space distribution is
\begin{equation}\label{FP}
\frac{\partial p(x,v,t)}{\partial t}=\beta_{J}p(x,v,t)-v\frac{\partial p(x,v,t)}{\partial x}+(\beta_{J}v+\sin x-\kappa)\frac{\partial p(x,v,t)}{\partial v}+D\frac{\partial^{2} p(x,v,t)}{\partial v^{2}}. \end{equation}
By virtue of It\^{o}'s formula and integration by parts, more accurate calculations for obtaining the Fokker-Planck equation can be found in the references \cite{YW,YZD}.

\begin{figure}[h]
\begin{center}
  \begin{minipage}{2.13in}
\leftline{(a)}
\includegraphics[width=2.13in]{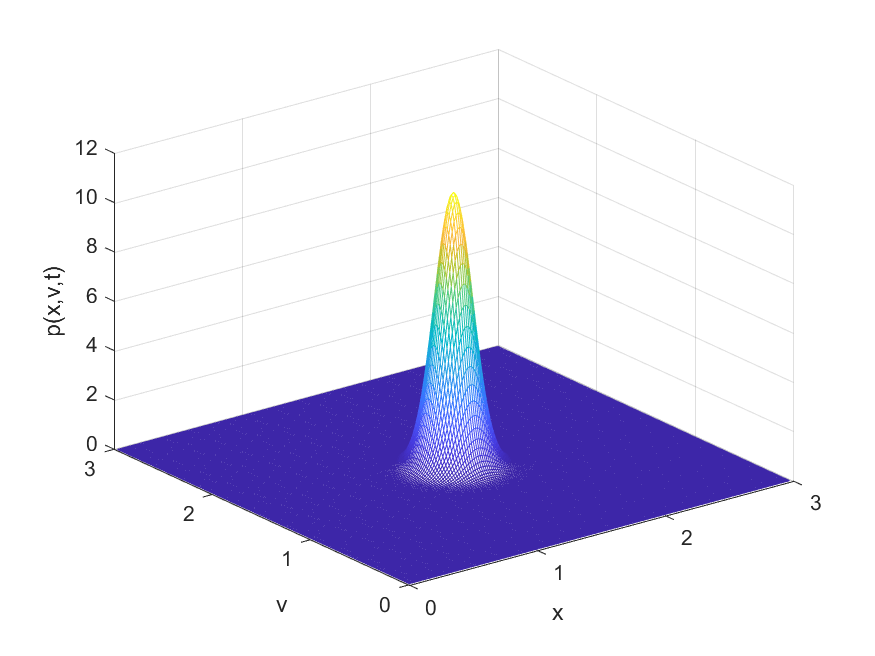}
\end{minipage}
\hfill
\begin{minipage}{2.13in}
\leftline{(b)}
\includegraphics[width=2.13in]{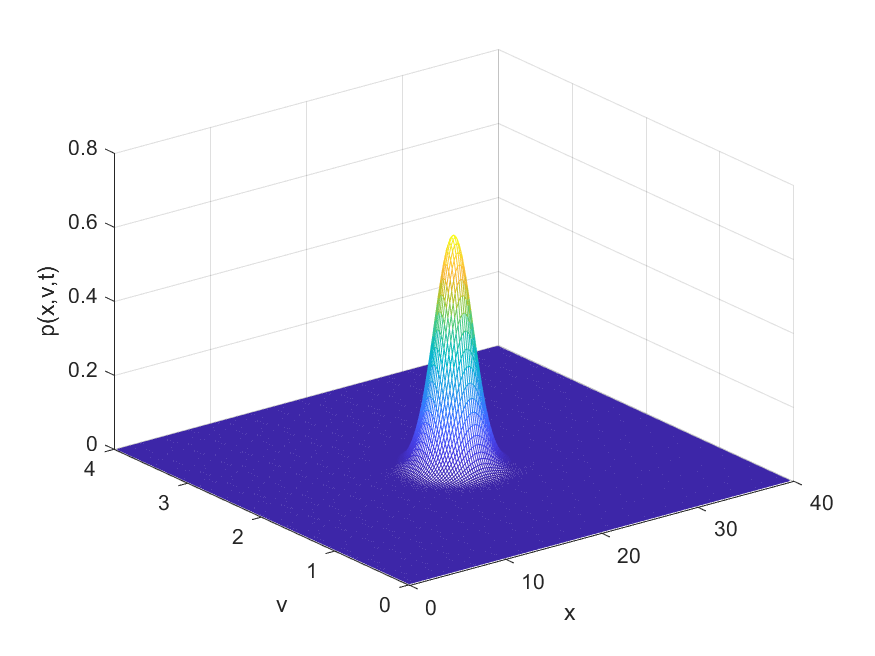}
\end{minipage}
\hfill
  \begin{minipage}{2.13in}
\leftline{(c)}
\includegraphics[width=2.13in]{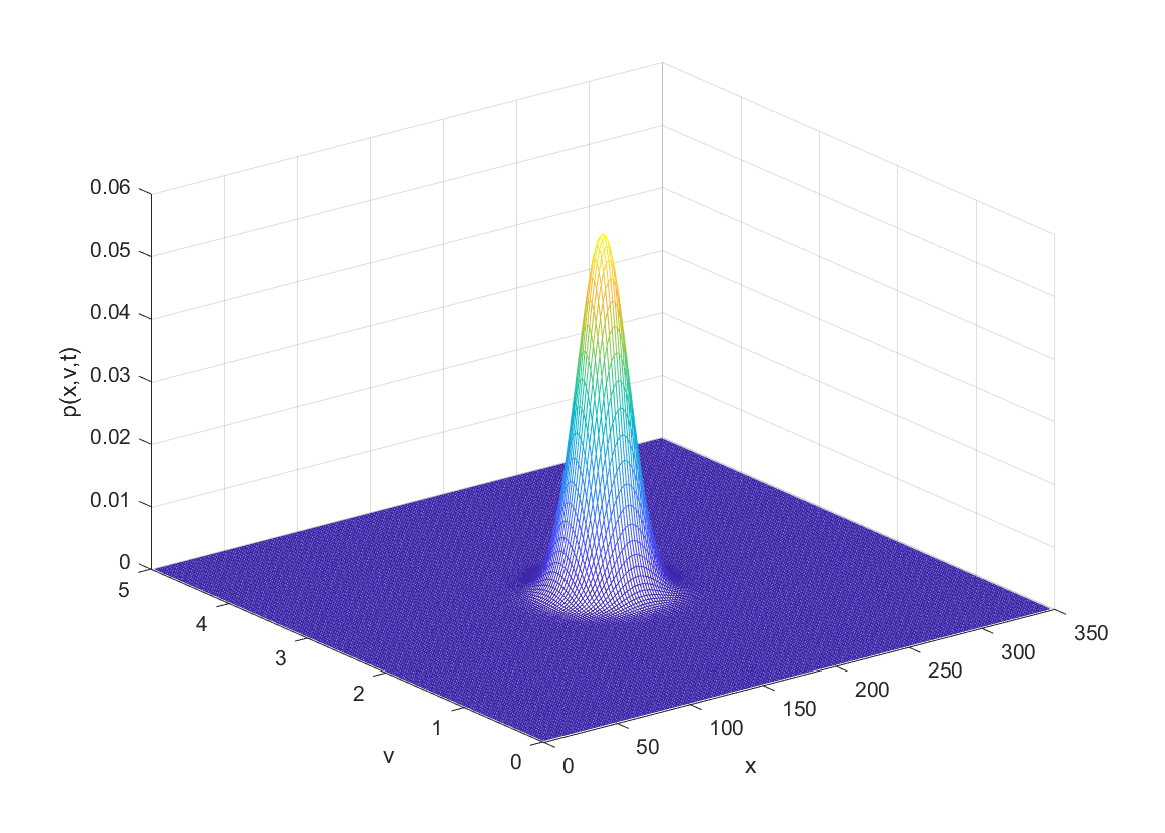}
\end{minipage}
\caption{ As the point $(x,v)$ evolves within the phase space, the three-dimensional simulations demonstrate the presence of a solitary peak in the distribution of $p(x,v,t)$ characterized by the Fokker-Planck equation \eqref{FP} with $\beta_J=1.2$ when considering a constant time value of $t=10$: (a) When $\kappa=0.1$, the peak's  height of probability density reaches a value of 12; (b) At the critical value of $\kappa=0.6965$, the peak of probability density experiences a decrease to 0.7; (c) When $\kappa=2$, the peak's height  of probability density further decreases to 0.06.
}\label{p}
\end{center}
\end{figure}

For analytical treatments it is much more convenient to work with phase-space
probabilities rather than individual stochastic trajectories. As the point $(x,v)$ evolves in the phase space, the three-dimensional  simulations reveal a single peak of the distribution of $p(x,v,t)$ for a fixed time $t=10$,  as shown in Figure \ref{p}. When the small parameter $\kappa=0.1$ is reached, Figure \ref{p} (a) indicates that  At the critical value $\kappa=0.6965$,
the peak of probability density actually decreases to 0.7 as clear from Figure \ref{p} (b).  In the case of $\kappa=2$, the peak's degree further decreases to 0.06, as exemplified in Figure \ref{p} (c). Overall, it was found that the peak of probability density is inversely proportional to the parameter $\kappa$.

\section{Summary and Outlook}\label{SO}
In the present work, the derivation of the Josephson equations for supercurrent and voltage was presented, leveraging the wave function of the superconducting tunnel junction and the time-dependent Schr\"{o}dinger equation that regulates its behavior. The equation governing the total junction current was analyzed in accordance with Kirchhoff's current law. By converting it to a dimensionless form, the impact of noise was introduced using the fluctuation-dissipation theorem, ultimately leading to the development of a resistively shunted superconducting tunnel junction system perturbed by thermal noise. The characteristics of the resistively shunted superconducting tunnel junction system were examined in the absence of parameters, revealing its Hamiltonian nature. The energy surface and phase portrait were simulated to gain further insights. In the presence of non-zero parameters, the system exhibited clockwise hysteresis in its typical I-V characteristic curve. For different parameter values, it was found that a bifurcation phenomenon affecting the planar limit cycle was possible. Furthermore, numerical simulations were conducted to visualize the stochastic dynamics of the system, particularly its response to thermal fluctuations. The Fokker-Planck equation corresponding to the resistively shunted superconducting tunnel junction system under thermal fluctuations was derived. Based on this equation, simulations were conducted to assess the probability density under specific parameter configurations.

There remains a sizable number of open problems in this direction. Among the significant challenges posed by the
experimental observations of charge noise generated by a mesoscopic conductor with a Josephson junction serving as
an on-chip detector, we note that the rate at which it switches out of its zero-voltage state is tied to the thermal escape
in the presence of non-Gaussian L\'evy noises \cite{CDSX,YB}. It would be extremely interesting to try to identify an effective non-local
Fokker-Planck equation, which facilitates the derivation of an explicit expression for the escape rate. Furthermore, the
non-zero third moment of current noise may lead to the formation of rate asymmetry. Exploring tunnel junctions
exhibiting L\'evy noise characteristics is an interesting question in its own right. Identifying these behaviors from the perspective of
dynamical systems, as presented herein, and more systematically examining the analysis of experimental
data and the optimization of detection circuits would constitute a particularly relevant task for future work.

\bigskip
\noindent\textbf{Data Availability}

Numerical algorithms source code that support the findings of this study are openly
available in GitHub, Ref. \cite{SY}.

\bigskip
\noindent\textbf{Declaration of competing interest}

No author associated with this paper has disclosed any potential or pertinent conflicts which
may be perceived to have impending conflict with this work.

\bigskip
\noindent\textbf{Acknowledgements}

The author gratefully acknowledges \emph{Syst\`emes Al\'eatoires Inhomog\`enes} organized by Institut Curie $\&$ Institut Henri Poincar\'e.
The author is happy to thank Professor Peter H\"{a}nggi for fruitful discussions on statistical mechanics.

%\section*{References}

%\bibliography{mybibfile}

\end{document}